\documentclass[a4paper,11pt,hyper]{JHEP3}
\usepackage{amsfonts,latexsym,graphicx,epsfig,amssymb,amsmath,mathrsfs}

\newcommand{\alt}{\hspace{0.3em}\raisebox{0.4ex}{$<$}\hspace{-0.75em}\raisebox{-.7ex}{$\sim$}\hspace{0.3em}}
\newcommand{\agt}{\hspace{0.3em}\raisebox{0.4ex}{$>$}\hspace{-0.75em}\raisebox{-.7ex}{$\sim$}\hspace{0.3em}}
\newcommand{\bch}{\bm{\chi}}
\newcommand{\sbch}{\sbm{\chi}}
\newcommand{\bm}[1]{\hbox{\boldmath{$#1$}}}
\newcommand{\sbm}[1]{\hbox{\boldmath{\scriptsize$#1$}}}
\newcommand{\ehzr}{|\, \Psi \rangle_{\bar{\zeta}^c}}

\newcommand{\SL}{\stackrel{\rm  s.l.}{\approx}}

\newcommand{\Seff}{S_{\rm eff}}

\newcommand{\Seffd}[1]{S'_{{\rm eff}(#1)}}

\newcommand{\dd}{d}

\title{Large gauge transformation, Soft theorem, and Infrared divergence in
inflationary spacetime}
\author{Takahiro Tanaka$^a$ and Yuko Urakawa$^b$
\\
a.~Department of Physics, Kyoto university,
 Kyoto, 606-8502, Japan\\
b.~Department of Physics and Astrophysics, Nagoya University, Chikusa,
Nagoya 464-8602, Japan
}

\abstract{It is widely known that the primordial curvature
perturbation $\zeta$ has several universal properties in the infrared
(IR) such as the soft theorem, which is also known as the consistency relation,
and the conservation in time. They are valid in rather general single clock models of inflation. It has been argued that these universal
properties are deeply related to the large gauge transformations in
inflationary spacetime. However, the invariance under the large gauge
transformations is not sufficient to show these IR properties. In this
paper, we show that the locality condition is crucial
to show the consistency relation and the conservation of $\zeta$. This
argument also can apply to an interacting system with the inflaton
and heavy fields which have arbitrary integer spins, including higher spin
fields, which may be motivated from string theory. We will also show
that the locality condition guarantees the cancellation of the IR
divergences in a certain class of variables whose correlation functions
resemble cosmologically observable quantities. }

\preprint{}

\maketitle

\begin{document}


\section{Introduction}
The presence of a symmetry helps us to systematically understand various
properties of a physical system. Understanding the physics based on the
symmetry may open up the possibility of understanding a specific example
based on a more universal property of physical systems which share
similar symmetry groups. It has been suggested that the infrared
(IR) properties of the primordial perturbations generated during
inflation are deeply related to the large gauge transformations. The
large gauge transformation is a local symmetry transformation which does
not approach the unity at the infinity of
spacetime~\cite{Harvey:1996ur, Avery:2015rga}. 

The large gauge transformations which play an important role in
cosmology are those defined at a time constant slicing and diverge at the
spatial infinity. In single field models of inflation, the invariance
under the dilatation and shear transformations, which are both large
gauge transformations, directly ensures the massless property of the
curvature perturbation $\zeta$ and the gravitational waves $\gamma_{ij}$. 
It is widely known that these massless fields $\zeta$ and $\gamma_{ij}$
have several universal properties in the IR, which are valid in rather
general single field models of inflation. First, at the tree-level computation, 
$\zeta$ and $\gamma_{ij}$ are non-linearly conserved in time in the IR
limit~\cite{WeinbergAd, WMLL, Starobinsky:1986fxa, Salopek:1990jq, Sasaki:1995aw, Sasaki:1998ug, MW03, LMS, LV05}. (The conservation of
$\zeta$ was discussed by including the radiative corrections of $\zeta$
in Refs.~\cite{SZ1210, ABG}.) Second, the influence from the soft modes
of $\zeta$ and $\gamma_{ij}$ on the hard modes are, rather generically, characterized
by the so-called consistency relation~\cite{Maldacena02, PM04}, which
is an example of the soft theorem in cosmology. It has
been suggested, e.g., in Refs.~\cite{WeinbergAd, HHK, Goldberger:2013rsa,
Berezhiani:2013ewa, TY_conservation, JY_CR}, that these universal properties are both consequences of
the invariance under the large gauge transformations. However, it was
also revealed that this invariance is not sufficient to derive the
consistency relation. In Refs.~\cite{Berezhiani:2013ewa,
TY_conservation, JY_CR}, the analyticity in the soft limit was
additionally imposed to derive the consistency relation.

The consistency relation has been understood as a relation between the
$n$-point correlation function with $n$ hard modes and  the
$(n+1)$-point correlation function with $n$ hard modes and one soft mode. In deriving
the consistency relation, usually, the validity of the perturbation 
theory is presumed. However, this is not trivially guaranteed, because
the massless fields $\zeta$ and $\gamma_{ij}$ are not screened in the
large scale limit and their radiative corrections can diverge due to
their unsuppressed IR contributions. Therefore, unless the IR divergence
is regularized without violating the dilatation invariance, the
consistency relation cannot be well-defined as a relation between the
correlation functions. Notice that introducing a naive IR cutoff which regularizes
the IR radiative corrections can violate the
dilatation invariance, which is a crucial property for the consistency
relation.

The goal of this paper is to clarify the relation between the large
gauge transformations and the above-mentioned universal properties of
$\zeta$ and $\gamma_{ij}$ in the IR, i.e., the soft theorem (also known
as the consistency relation), the conservation in time, and the possible
appearance of the IR divergence. In particular, along the line of the argument in
Refs.~\cite{Berezhiani:2013ewa,TY_conservation, JY_CR}, we will clarify
the physical meaning of the analyticity in the soft limit, which is
needed to derive the consistency relation. In Ref.~\cite{Berezhiani:2013ewa}, it was argued that
this condition is related to the locality of theory. In this paper, we
scrutinize this argument, clarifying what we need to require as the
locality condition, because the locality has a broad meaning. We also
discuss what the consistency relation actually describes, taking into
account the possibility that the perturbative prediction can be spoiled by
the IR divergence.

We will also show that the large gauge transformations play a crucial
role also in discussing the radiative corrections from massive
fields. Being motivated by string theory, which may predict the presence
of higher spin fields in the four dimensional theory obtained after
compactification, we consider massive fields with arbitrary integer 
spins. After deriving the consistency relation for the hard modes of the
massive fields with non-zero spins, we discuss the condition that the curvature perturbation
$\zeta$ stops evolving in time in the soft limit under the influence of
the radiative corrections from the massive fields. This is a
generalization of our previous study \cite{TY_conservation}, where we considered the 
radiative corrections from a massive scalar field, to a massive field
with a general integer spin. As was argued in Ref.~\cite{NJ15}, exploring an
imprint of the higher spin fields in the primordial non-Gaussianity of
$\zeta$ may provide a unique probe of string theory. (See also
Refs.~\cite{CW09, Noumi:2012vr, MM15, Lee:2016vti}.) Our study provides the
conditions for the absence of such an imprint made after the Hubble 
crossing time of the comoving scale of our interest during inflation.

Recently, the relation among the large gauge transformations, the soft
theorem, and the IR divergence has been studied intensively for gauge
theories in an asymptotically flat spacetime~\cite{Strominger:2013jfa,
He:2014laa, Strominger:2014pwa}. (For a review, see Ref.~\cite{Strominger:2017zoo}.) In Refs.~\cite{Strominger:2013jfa}, it was shown
that Weinberg's soft theorem~\cite{WeinbergST}, which describes the influence of the soft
photon and graviton, can be obtained as a Ward-Takahashi identity of the
asymptotic symmetries at the null infinity. (The soft theorem for massless higher spin fields in an asymptotically flat spacetime was discussed in Ref.~\cite{Campoleoni:2017mbt}.) More recently, (the
cancellation of) the IR divergence was discussed, using the Noether
charge of the asymptotic symmetries~\cite{Kapec:2017tkm}. At first sight, the IR structures for the gauge
fields in the asymptotically flat spacetime have similar properties to
those for the primordial perturbations $\zeta$ and $\gamma_{ij}$ in
cosmology. We give a closer look at this apparent similarity.

This paper is organized as follows. In Sec.~\ref{Sec:LGT}, considering single field models of inflation, we discuss the relation between the large gauge transformation and the consistency relation for
$\zeta$. Here, we seek for a deeper understanding about the locality
condition as the necessary condition to derive the consistency relation for $\zeta$. In Sec.~\ref{Sec:spin}, we will show
that the discussion in Sec.~\ref{Sec:LGT} can be straightforwardly
extended to an interacting system composed of the inflaton and the
massive fields with non-zero spins. We also show that the locality
condition implies both the consistency relation for the hard modes of
the massive fields and the conservation of $\zeta$. In Sec.~\ref{Sec:IR},
we show that the locality condition also ensures the cancellation of the
IR divergence for a certain class of variables. In Sec.~\ref{Sec:GW},
we briefly show that the discussion for the soft graviton proceeds
almost in parallel to the one for the soft modes of $\zeta$, discussed in
Sec.~\ref{Sec:LGT}-\ref{Sec:IR}. In Sec.~\ref{Sec:Conclusion}, after
summarizing our results, we discuss a similarity and an apparent
difference between the IR properties for the gauge fields in the
asymptotically flat spacetime and those for the primordial perturbations.

\section{Asymptotic symmetry and soft theorem}  \label{Sec:LGT}
In this section, we discuss the relation between the large gauge
transformations and the soft theorem in cosmology, also known as the
consistency relation. We clarify the condition that derives the consistency
relation and discuss its physical meaning. Before we start our
discussion, we clarify what we mean by ``soft'' and
``hard.'' The soft modes mean the modes with $k/aH \to 0$ but
$\bm{k} \neq 0$, and the hard modes mean the remaining modes, including the super Hubble
modes with $k/aH \alt 1$ but excluding the limit $k/aH \to 0$. By contrast, we describe
the wave vector of a longer mode and that of a shorter mode as 
$\bm{k}_L$ and $\bm{k}_S$, respectively, simply based on the ratio between these wavenumbers,
i.e., $k_L/k_S \ll 1$. As we will discuss in Sec.~\ref{Sec:IR}, in the
limit $k/aH \ll 1$, a perturbative expansion can break down in
computing some quantity and taking this limit requires a careful treatment.

\subsection{Large gauge transformations}
Likewise in the discussion about the soft photons and gravitons in the
asymptotically flat spacetime, a large gauge transformation plays a
crucial role for a clear understanding about the soft modes of $\zeta$
and $\gamma_{ij}$ in an inflationary spacetime. In line with
Refs.~\cite{Harvey:1996ur, Avery:2015rga}, we define the large gauge
transformation as follows. A local symmetry denotes a symmetry under a
transformation which is parametrized by a spacetime dependent function,
while a global symmetry denotes a symmetry under a transformation by a
spacetime independent function.

Among local symmetry transformations, it is important to make a
distinction between small gauge transformation and large gauge
transformations. The former becomes the identity at the infinity 
and the latter does not.
In Refs.~\cite{Strominger:2013jfa, Strominger:2017zoo}, it was shown that the soft theorem
for the photons and the gravitons in the asymptotically flat spacetime
can be derived from the Ward-Takahashi identities for large gauge
transformations which do not vanish on ${\cal J}^{\pm}$.  

\subsubsection{Dilatation as a large gauge transformation}  \label{SSSec:dilatation}
First, let us clarify the prescription we adopt. In this paper, we use the ADM form of the line element: 
\begin{align}
 d s^2 = - N^2 d t^2  + h_{ij} (d x^i + N^i d t) (d x^j + N^j d t)~,
 \label{Exp:ADMmetric}
\end{align}
where we introduced the lapse function $N$, the
shift vector $N^i$, and the spatial metric $h_{ij}$. We determine the
time slicing, employing the uniform field gauge: 
\begin{align}
 & \delta \phi=0\,.   \label{GCphi}
\end{align} 
We express the spatial metric $h_{ij}$ as
\begin{align}
 & h_{ij}= a^2 e^{2\zeta} \left[ e^{\gamma} \right]_{ij}\,,
\end{align}
where $\gamma_{ij}$ is set to traceless. As spatial gauge conditions,
we impose 
\begin{align}
 & \partial^i \gamma_{ij} =0\,. \label{GCT}
\end{align}

To discuss the soft modes of the primordial perturbations in the
spatially flat FRW background, we consider the large gauge
transformations, which do not vanish at the spatial infinity on a time
constant surface. This large gauge transformation was first discussed in
the context of cosmology by Weinberg in Ref.~\cite{WeinbergAd}. In the
unitary gauge, where the fluctuation of the inflaton vanishes, we
consider, in particular, the dilatation:
\begin{align}
 & x^i \to e^s x^i\,,
\end{align} 
where $s$ is a constant parameter. Under the dilatation, the curvature
perturbation $\zeta$ transforms as
\begin{align}
 & \zeta(t,\, \bm{x}) \to \zeta_s(t,\, \bm{x})= \zeta(t,\, e^{-s}
 \bm{x}) -s\,. \label{Exp:zetaD}
\end{align}
The change of $\zeta$ is given by
\begin{align}
 & \Delta_s \zeta(t,\, \bm{x}) = - s (1 + \bm{x} \cdot \partial_{\sbm{x}} \zeta(t, \, \bm{x}))
 + {\cal O}(s^2)\,. 
\end{align}

The classical action in a diffeomorphism (Diff)
invariant theory remains invariant under the transformation of $\zeta$
given in Eq.~(\ref{Exp:zetaD}). As one may expect from the fact that the
dilatation shifts $\zeta$ by $-s$, the dilatation
invariance is related to the massless property of $\zeta$, which implies
that $\zeta$ is conserved at large scales in single clock inflation.

\subsubsection{Two different prescriptions of dilatation}
The dilatation invariance may be somehow confusing, because it also
appears as a part of the de Sitter invariance by changing the time coordinate
simultaneously. The Killing vector which corresponds to this transformation is given by
$$
- \eta \partial_\eta - x^i \partial_i\,,
$$
where $\eta$ denotes the conformal time. The dilatation symmetry in de
Sitter group states
that the time shift can be compensated by the scale
transformation. Since inflation has to end at some point, the time
translation symmetry needs to be broken in the context of inflationary scenario.

The time translation symmetry is broken, when the physical frequency
$\omega_{ph}$ becomes well below $\Lambda_b$ with $\Lambda_b^4 \equiv \dot{\phi}^2$.
In the effective field theory of inflation~\cite{EFToI}, the Goldstone mode, the pion
$\pi$, is introduced to restore the invariance under the 
time reparametrization  
\begin{align}
 & t \to t + \xi\,, \qquad \pi \to \pi - \xi
\end{align}
in the symmetry breaking phase. With this construction,
the pion Lagrangian non-linearly preserves the
invariance since $\pi$ appears only in the combination $t+ \pi$. 
Through the
coupling with the metric perturbations, the pion acquires the mass of
$ m_\pi = {\cal O}(\sqrt{\varepsilon_1} H)$.

The relation between the dilatation discussed in
Sec.~\ref{SSSec:dilatation} and the one discussed here is somewhat
puzzling. As we will discuss in the following section, the former is
preserved in an arbitrary quasi FRW spacetime, while the latter is a
part of the de Sitter symmetry and is broken below the symmetry breaking
scale $\Lambda_b$. Related to this point, preserving the former
dilatation invariance directly ensures that $\zeta$ should be massless. On the other hand, there is no
simple argument which shows the massless property of $\zeta$ in the
latter prescription, where $\zeta$ is related to the Goldstone mode $\pi$ as
$\zeta = - H \pi$. Related to this point,
recall that the pion acquires the mass $m_\pi$ through the coupling with
the metric perturbation for $\omega_{ph} \alt m_\pi$~\footnote{Recall that
the Goldstone mode for a global symmetry is not necessarily
massless in a Lorentz violating background.}. Therefore, in the
regime where $\zeta$ approaches a constant value, the pion is no
longer massless. The curvature perturbation is sometimes said to be a
Goldstone mode, since $\pi= - \zeta/H$ is the Goldstone mode. This
statement can cause a confusion, because it may
sound as if $\zeta$ preserves the shift symmetry, being massless, because it is a Goldstone boson associated with the breaking of the de Sitter symmetry.

In the following, by the dilatation, we mean the former one, which is a
spatial coordinate transformation without the time coordinate
change. Considering this dilatation, we discuss the IR behaviour of
$\zeta$ such as the consistency relation and the conservation in time. We 
will emphasize that the invariance under this dilatation, which is a part of the large gauge
transformations, is preserved as well in the quantized system and
therefore there is no spontaneous symmetry breaking in this prescription.

\subsection{Dilatation invariance and Noether charge}
In this subsection, we discuss several implications of the dilatation
invariance in single clock inflation. Following Ref.~\cite{HHK}, we
define the Noether charge for the dilatation as 
\begin{align}
 & Q_\zeta \equiv \frac{1}{2} \int d^3 \bm{x} \left[  \Delta_s \zeta(t,\,
 \bm{x}) \pi_\zeta (t,\,  \bm{x}) + \pi_\zeta(t, \, \bm{x}) \Delta_s
 \zeta(t,\, \bm{x}) \right]\,,  \label{Exp:QzetaP}
\end{align}
where $\pi_\zeta$ denotes the conjugate momentum of $\zeta$, which
satisfies 
\begin{align}
 & \left[ \zeta(t,\, \bm{x}),\, \pi_\zeta (t,\, \bm{y}) \right] = i
 \delta (\bm{x} - \bm{y})\,.  \label{Exp:commutation}
\end{align} 
Since the Hamiltonian for $\zeta$ is invariant under the dilatation, we
obtain
\begin{align}
 & \left[ Q_\zeta,\, H \right] = 0\,,
\end{align}
which implies that $Q_\zeta$ is independent of time. The Noether charge
is a generator of the dilatation transformation and satisfies
\begin{align}
 & \left[ Q_\zeta,\, \zeta(x) \right] = - i \Delta_s \zeta(x) \,. \label{Exp:ComQ}
\end{align}

Using the Fourier components of the fields\footnote{We use the convention of the Fourier transformation:
$$
 f(\bm{x}) = \int \frac{d^3 \bm{k}}{(2 \pi)^3} e^{i \sbm{k} \cdot
 \sbm{x}} \hat{f}(\bm{k}) \,, \qquad 
 \hat{f}(\bm{k}) = \int d^3 \bm{x} e^{- i \sbm{k} \cdot \sbm{x}} f(\bm{x})\,.
$$
Here, the commutation relation for the Fourier modes of $\zeta$ and $\pi_\zeta$ is given by
$ [\zeta_{\sbm{k}},\, \pi_{\zeta\, \sbm{k}'}] = i (2 \pi)^3 \delta(\bm{k} + \bm{k}')$.
} 
we can rewrite the Noether charge $Q_\zeta$ as
\begin{align}
 & Q_\zeta = - s \pi_{\zeta, \sbm{k}=0} - \frac{s}{2} \int \frac{d^3
 \bm{k}}{(2 \pi)^3} \, \left\{  \zeta_{\sbm{k}},\, \bm{k} \cdot \partial_{\sbm{k}} \pi_{\zeta,\,
 - \sbm{k}} \right\} + {\cal O}(s^2)\,.   \label{Exp:QzetaF}
\end{align}
In performing the Fourier transformation, we did not
drop the surface term. Therefore, the charge $Q_\zeta$ given in
Eq.~(\ref{Exp:QzetaP}) is identical to the one given in
Eq.~(\ref{Exp:QzetaF}). The first term of $Q_\zeta$ only operates 
on the $\bm{k}=0$ mode. 
The Noether charge $Q_\zeta$ can diverge due to the IR
modes, because it is an integral over the infinite spatial volume. In
the following, we neglect higher order terms of ${\cal O}(s^2)$.

Equation (\ref{Exp:QzetaF}) gives the non-perturbative definition of the
Noether charge for the dilatation. Since an explicit computation usually
relies on perturbative expansion in the interaction picture, one may
want to introduce the Noether charge by using the fields in the
interaction picture as\footnote{Notice that since
the free Hamiltonian $H_0$ is not invariant under the dilatation,
i.e.,
\begin{align}
 & \left[ H_0,\, Q_\zeta^I \right] \neq 0\,,
\end{align}
in contrast to $Q_\zeta$, which is time independent, the
``charge'' defined in the interaction picture $Q_\zeta^I$ varies in
time. 
}
\begin{align}
 & Q_\zeta^I \equiv \frac{1}{2} \int d^3 \bm{x} \left[  \Delta_s \zeta^I(t,\,
 \bm{x}) \pi^I_\zeta (t,\,  \bm{x}) + \pi^I_\zeta(t, \, \bm{x}) \Delta_s
 \zeta^I(t,\, \bm{x}) \right]\,,  \label{Exp:QzetaPI}
\end{align}
where $\Delta_s \zeta^I(t,\, \bm{x})$ is given by
\begin{align}
 & \Delta_s \zeta^I(t,\, \bm{x}) = - s (1 + \bm{x} \cdot
 \partial_{\sbm{x}} \zeta^I(t, \, \bm{x})) \,. 
\end{align}

When we perform the dilatation transformation in the interaction picture
or after the perturbative transformation, there is one caveat which
should be kept in mind. Changing order of performing the finite
dilatation transformation and performing the perturbative expansion
leads us to a different answer. In order words, performing the
dilatation in the interaction picture (with a use of $Q_\zeta^I$)
and performing the dilatation in the Heisenberg picture (with $Q_\zeta$)
are different, while this discrepancy disappears, when we
consider the infinitesimal transformation generated by 
$d Q_\zeta/d s|_{s \to 0}$. A lesson from here is that for a legitimate
prescription, the dilatation transformation should be performed in the non-perturbative Heisenberg
picture. A more detailed discussion can be found in Appendix~\ref{Ap:NC}.

\subsection{Revisiting consistency relation}
In this subsection, we discuss the condition(s) to derive the
consistency relation of $\zeta$. We will see that while the dilatation
invariance plays a crucial role, it is neither sufficient nor even 
necessary to derive the
consistency relation.

\subsubsection{Condition for consistency relation} \label{SSSec:ConditionCR}
The consistency relation for $\zeta$ was first derived by Maldacena
from an explicit computation of the bi-spectrum of $\zeta$~\cite{Maldacena02}.
Afterwards, the consistency relation was derived in a
rather general setup of single field models of inflation~\cite{PM04}, while several
examples that do not satisfy the consistency relation were also
reported~\cite{NFS, CNFS, Berezhiani:2014kga} (see also
Refs.~\cite{Dimastrogiovanni:2014ina, Dimastrogiovanni:2015pla}).

In Ref.~\cite{Maldacena02}, it was argued that the bi-spectrum in the
squeezed limit, which describes the correlation between the long mode $\bm{k}_L$
and short mode $\bm{k}_S$, can be computed by considering the influence of the long
mode on the short mode. This influence can be described as the dilatation
$\bm{k}_S \to e^{-\zeta_{\sbm{k}_L}} \bm{k}_S$. While $\zeta$ is not
invariant under the dilatation transformation\footnote{The curvature
perturbation $\zeta$ is invariant under the small gauge transformations
but not under the large gauge transformations.}, we can construct another
variable which is invariant under the dilatation
transformation~\cite{IRgauge_L, IRgauge}. In Ref.~\cite{IRNG}, we showed
that the leading contributions to the bi-spectrum of such an invariant variable in
the squeezed limit are canceled, choosing the adiabatic vacuum. In this computation, the squeezed bi-spectrum of
$\zeta$ satisfies Maldacena's consistency relation, which is the key to
ensure the cancellation.

The consistency relation does not hold for an arbitrary quantum state
even in single field models. In Refs.~\cite{Goldberger:2013rsa,
Berezhiani:2013ewa, JY_CR}, it has been directly and indirectly suggested that the dilatation
invariance of the quantum state:
\begin{align}
 & Q_\zeta | \Psi \rangle = 0 \label{Cond:Inv}
\end{align}
is crucial to derive the consistency relation. In
Ref.~\cite{Goldberger:2013rsa}, it was shown that the invariance of the
1PI effective action for $\zeta$ under the dilatation leads to a set 
of identities which relate the $n$-point function and the $(n+1)$-point
function for $\zeta$. What we directly obtain from these identities,
which are the so called Ward-Takahashi (WT) identity, is ``the
consistency relation,'' which relates the $n$-point function with
$n$ hard modes to the $(n+1)$-point function with
the homogeneous mode $\zeta_{\sbm{k}=0}$ 
in addition to the $n$ hard modes. The only difference
from the consistency relation is in that the inserted mode is not the soft mode $\zeta_{\sbm{k}}$ with
$\bm{k} \neq 0$.

Along this line, in Refs.~\cite{JY_CR, Kundu:2014gxa, Kundu:2015xta}, ``the consistency
relation'' was derived starting with the dilatation invariance of the wave function in the
$\zeta$ representation, 
$\Psi[\zeta] \equiv \langle \zeta| \Psi \rangle$, where 
$|\, \zeta \rangle$ is the normalized eigenstate of $\zeta$, i.e.,
\begin{align}
 & \Psi[\zeta_s(t,\, \bm{x})] = \Psi[\zeta(t,\, \bm{x})]\,, \label{Exp:InvPsi}
\end{align}
which directly follows from the invariance of the quantum state under
the dilatation (\ref{Cond:Inv})\footnote{For instance, at 
${\cal O}(s)$, Eq.~(\ref{Exp:InvPsi}) gives 
$$
0=\int d^3\bm{x}\, \Delta_s \zeta(t,\, \bm{x}) \frac{\delta}{\delta
\zeta(t,\, \bm{x})} \psi[\zeta]\,, 
$$
which can be obtained from 
$0 = \langle \zeta(x) | Q_\zeta | \psi \rangle$ by inserting the
expression of the Noether charge, given in Eq.~(\ref{Exp:QzetaP}). Here,
we used 
$\langle \zeta| \pi_\zeta | \psi \rangle \propto \delta  \psi[\zeta]/\delta \zeta$. 
}. 
If and only if the WT identity which describes the insertion of $\zeta_{\sbm{k}=0}$
can be extended to the relation which describes the insertion of the soft mode
$\zeta_{\sbm{k}}$ with $\bm{k} \neq 0$, we obtain the consistency relation:
\begin{align}
 & \lim_{k_n \to 0} \frac{{\cal C}^{(n)}(\{\bm{k}_i\}_n)}{P(k_n)} = -
 \left( \sum_{i=2}^{n-1} \bm{k}_i \cdot 
 \partial_{\sbm{k}_i} + 3(n-2) \right) {\cal
 C}^{(n-1)}(\{\bm{k}_i\}_{n-1})\,, \label{Exp:CRd}
\end{align}
where ${\cal C}^{(n)}$ denotes the $n$-point function of $\zeta$ with the momentum conservation factor
$$
(2\pi)^3 \delta \left( \sum_{i=1}^{n} \bm{k}_i \right)
$$
removed. Since we additionally need to impose that this extension is
possible, requesting the dilatation invariance of the quantum state
(\ref{Cond:Inv}) is not enough to derive the consistency
relation.

Now, the question is ``What is the physical meaning of the additional
condition that allows the WT identity to be smoothly extended to the
consistency relation, which describes the insertion of the soft mode?''
This issue was first addressed in Ref.~\cite{Berezhiani:2013ewa}, where
it was argued that this condition is related to the locality of the theory. Even if the
original theory is local, the Lagrangian density for $\zeta$ in the unitary gauge becomes 
non-local due to the presence of the Lagrange multipliers,
the lapse function and the shift vector, which are given by solving the
elliptic equations. For instance, at the linear order in perturbation,
the shift vector $N_i$ includes a contribution given by 
$$
 \partial_i N_i \supset \varepsilon \dot{\zeta} \,,
$$
which introduces non-local interaction vertices. In the standard slow-roll inflation, 
$\dot{\zeta}_{\sbm{k}}$ is suppressed in the limit $\bm{k} \to 0$ as 
$\dot{\zeta}_{\sbm{k}} = {\cal O}(k^p \zeta_{\sbm{k}})$ with $p \geq 1$. However, in the absence of this suppression, 
the coefficients of the non-local interaction vertices in the Fourier
space can be singular in the limit $\bm{k} \to 0$. Then, $N_i$ should be
determined discontinuously at $\bm{k} = 0$ in order to avoid the singular behaviour.

Along this line, our purpose of this section is to sharpen the relation
between the condition for the locality and the
condition of being able to extend the ``consistency relation'' with the insertion of the
homogeneous mode to the consistency relation with the soft mode $\bm{k} \neq 0$, clarifying the physical meaning of the
condition. As was mentioned above, the Lagrangian density for $\zeta$
(and also for the gravitational waves) is non-local in the
sense that the Lagrangian density cannot be solely determined by the
dynamical fields at each spacetime point. For a classical theory, the
condition for the smooth extension of the homogeneous mode $\bm{k}=0$ to
the soft modes $\bm{k} \neq 0$ with a suitable fall-off
at the spatial infinity is nothing but the one to pick up the Weinberg's adiabatic mode~\cite{WeinbergAd}. We
elaborate the physical meaning of this condition for a quantum theory by
using the Noether charge $Q_\zeta$. As will be discussed in the next section, with the use
of the Noether charge, a generalization to the case with higher spin fields
proceeds straightforwardly.

\subsubsection{Conditions on the homogeneous mode and hard modes}
\label{SSSec:NandNP}

As was mentioned in Sec.~\ref{SSSec:ConditionCR}, the validity of the consistency
relation is deeply related to the invariance of the wave
function or the effective action under the dilatation, which is a large
gauge transformation. Maldacena derived the consistency relation for the tree-level
bi-spectrum in the squeezed limit, i.e., (\ref{Exp:CRd}) with $n=3$,
choosing the adiabatic vacuum (or the Euclidean
vacuum)~\cite{Maldacena02}. This vacuum also can be defined
non-perturbatively by requesting the regularity of correlation functions
in the limit $t \to - \infty(1 \pm i \epsilon)$. Here, the time path is
rotated towards the imaginary axis in the distant past. This serves one
of the examples of the quantum state $| \Psi \rangle$, which preserves
the dilatation invariance, since this definition does not artificially
introduce any specific scale. At perturbative level, a correlation function for the Euclidean vacuum can
be calculated by adopting the $i \epsilon$ prescription. Here, the
(free) mode function should be chosen to be the one for the adiabatic
vacuum (a.k.a, the Bunch-Davies vacuum in de Sitter
limit)~\cite{SRV2}.

In order to study the condition for the dilatation invariance of the
quantum state (\ref{Cond:Inv}), we decompose the wave function in terms of the
eigenstates of the spatial average of $\zeta$ all over a time constant slicing,
e.g., at the end of inflation, 
\begin{align}
 & \bar{\zeta} \equiv \frac{\int d^3 \bm{x} \zeta(\bm{x})}{\int d^3
 \bm{x}} \,, 
\end{align}
as
\begin{align}
 & |\Psi \rangle = \int d \bar{\zeta}^c\, |\psi(\bar{\zeta}^c)|\, |\,
 \bar{\zeta}^c\, \rangle \ehzr \label{Exp:psiz00}\,.
\end{align}
The eigenstate $ |\, \bar{\zeta}^c\, \rangle$ satisfies
\begin{align}
 & \bar{\zeta}  |\, \bar{\zeta}^c\, \rangle = \bar{\zeta}^c  |\, \bar{\zeta}^c\, \rangle\,,
\end{align}
where $\bar{\zeta}^c$ is a c-number eigenvalue. In order to distinguish the eigenvalues of $\bar{\zeta}$, which are
$c$-numbers, from the operator $\bar{\zeta}$, we put the index $c$ on the
eigenvalues. In Eq.~(\ref{Exp:psiz00}), we factorized the wave function of $\bar{\zeta}$, $\psi(\bar{\zeta}^c)$, from 
$\langle \bar{\zeta}^c\,| \Psi \rangle$, while absorbing the phase into $\ehzr$, as
\begin{align}
 & \langle \bar{\zeta}^c\,| \Psi \rangle = |\psi(\bar{\zeta}^c)|\, \ehzr\,.  \label{Def:ehzr}
\end{align}
From the normalization condition for $\ehzr$, the amplitude of the 
wave function $ |\psi(\bar{\zeta}^c)| $ is unambiguously defined. 
Since the quantum state $|\,\Psi \rangle$ also includes the
inhomogeneous modes with $\bm{k} \neq 0$, 
$\langle \bar{\zeta}^c\,| \Psi \rangle $ should be understood as a
vector in infinite dimensional Hilbert space. We express the normalized quantum state for all the modes with $\bm{k} \neq 0$ obtained by the
projection of $|\,\Psi \rangle$ to the eigen state $|\, \bar{\zeta}^c \rangle$ as
$\ehzr$. Using Eq.~(\ref{Exp:ComQ}), we obtain
\begin{align}
 & \left[ i Q_\zeta,\, \bar{\zeta} \right] = - s\,. 
\end{align} 
Since an operation of $e^{i Q_\zeta}$ shifts the eigenvalue of
$\bar{\zeta}$ by $s$, i.e.,
\begin{align}
 & e^{i Q_\zeta} |\, \bar{\zeta}^c \rangle = |\, \bar{\zeta}^c + s
 \rangle \,, 
\end{align}
we obtain
\begin{align}
 & i Q_\zeta |\, \bar{\zeta}^c \rangle = s \frac{\partial}{\partial
 \bar{\zeta}^c} |\, \bar{\zeta}^c \rangle  \label{Exp:Q0}\,.
\end{align}

Using this condition, we can express the dilatation invariance condition
for the quantum state $|\, \Psi \rangle$, (\ref{Cond:Inv}), as
\begin{align}
 & 0 = \int d \bar{\zeta}^c \left[- s \frac{\partial |\psi(\bar{\zeta}^c)|
 }{\partial \bar{\zeta}^c} |\, \bar{\zeta}^c \rangle \ehzr + |\psi
 (\bar{\zeta}^c)|\,  |\, \bar{\zeta}^c \rangle \left( i Q_\zeta - s \frac{\partial}{\partial
 \bar{\zeta}^c}  \right) \ehzr  \right]\,. \label{Cond:Inv2}
\end{align}
Operating $\langle {\bar{\zeta}^{c'}}\,|$ on
Eq.~(\ref{Cond:Inv2}), the real and imaginary parts, respectively, give 
\begin{align}
 &  \frac{\partial}{ \partial \bar{\zeta}^c}\, |\psi(\bar{\zeta}^c)| = 0\,, \label{Cond:h0}
\end{align} 
and  
\begin{align}
 & i Q_\zeta  \ehzr =  s \frac{\partial}{\partial
 \bar{\zeta}^c} \ehzr \,. \label{Cond:inh0}
\end{align}
The first condition (\ref{Cond:h0}) requires that the amplitude of the
wave function $\psi(\bar{\zeta}^c)$, which represents the probability
distribution of $\bar{\zeta}$, should be flat in the direction of $\bar{\zeta}$ in the Hilbert
space. The condition (\ref{Cond:h0}) is satisfied, e.g., for a Gaussian wave function  
whose variance blows up in the limit of the homogeneous mode,  
which is the case for a nearly scale invariant power spectrum. 
The condition (\ref{Cond:h0}) requires that
the probability distribution should be non-perturbatively flat in the direction
of the homogeneous mode $\bar{\zeta}$. The second condition may be a little more
non-trivial. The quantum state of the inhomogeneous modes generically
changes under a variation of the homogeneous mode $\bar{\zeta}$. The
condition (\ref{Cond:inh0}) restricts how the quantum state of the inhomogeneous modes
should respond to the change of the homogeneous mode, which is
expressed by the operation of $Q_\zeta$.

Since the consistency relation describes an insertion of an
inhomogeneous mode with $\bm{k} \neq 0$ that suitably falls off at the
spatial infinity, we need to extend the above argument to
inhomogeneous modes. For this purpose, we introduce the generator of a
spatial dependent dilatation given by 
\begin{align}
 & Q^W_\zeta (\bm{x}) \equiv \frac{1}{2} \int d^3 \bm{x}' W(\bm{x}' - \bm{x}) \left[  \Delta_s \zeta(t,\,
 \bm{x}') \pi_\zeta (t,\,  \bm{x}') + \pi_\zeta(t, \, \bm{x}') \Delta_s
 \zeta(t,\, \bm{x}') \right]\,,
\end{align}
where $W(\bm{x})$ denotes a smooth window function which is normalized as
$\int d^3 \bm{x}\, W(\bm{x}) = 1$ and which vanishes at $|\bm{x}| \agt L$.
Here, we set $L$ to be of order of $1/k_L$. The generator $Q^W_\zeta (\bm{x})$ induces the dilatation
only for the fields at $\bm{x}'$ with $|\bm{x} - \bm{x}'| \alt L$. The Noether charge (\ref{Exp:QzetaP})
is defined by the integral all over the time constant slicing and hence the integral
does not converge in general. The introduction of the window function
can make the integral converge. Since the domain of integration is
finite, the generator $Q_\zeta^W(\bm{x})$ depends on where we choose the center
of the integral domain, $\bm{x}$. Performing the Fourier transformation, we obtain
\begin{align}
 & Q_\zeta^W (\bm{k}_L) = - s_{- \sbm{k}_L} \left[
 \pi_{\sbm{k}_L} + \frac{1}{2} \int \frac{d^3 \bm{k}}{(2 \pi)^3}
  \left\{ \zeta_{\sbm{k}},\, \bm{k} \cdot \partial_{\sbm{k}} \pi_{\sbm{k}_L - \sbm{k}} \right\}
 \right] + {\cal O}(s^2)\,,  
\end{align}
where $\hat{W}(\bm{k})$ denotes the Fourier mode of the window function,
which is normalized as 
$$
 \lim_{\sbm{k} \to 0} \hat{W}(\bm{k}) =1\,,
$$
and vanishes for $k \gg 1/L$. Here, we introduced
\begin{equation}
 s_{\sbm{k}}\equiv  s\hat W( \bm{k})\,.
\end{equation}

We focus on the field within a large volume of ${\cal O}(L_c^3)$. 
We set $L_c$ to be much larger than all the 
wavelengths, i.e., $L_c \gg 1/k_L, 1/k_S$, and we will send $L_c \to \infty$ 
after our computation. 
Next, we introduce a smeared field in momentum space defined by 
\begin{equation}
  \tilde \zeta_{\sbm{k}_L}\equiv
  L_c^3 \int d^3\bm{k}' \,\hat{\cal W}(\bm{k}_L-\bm{k}') \,\zeta_{\sbm{k}'}\,,
\end{equation}
with 
\begin{equation}
   \hat{\cal W}(\bm{k})=\prod_{i=1}^3\theta\left[(2L_c)^{-1}\!\! -|k_i|\right]\,.
\end{equation}
$\tilde \zeta_{\sbm{k}_L}$ describes a collective mode with the
representative wavenumber $\bm{k}_L$ and when we evaluate
$\tilde{\zeta}$ in the position space by performing the inverse Fourier
transformation, it decays outside the local volume of ${\cal O}(L_c^3)$. The
fluctuations outside the local volume can be described as the modes
which are orthogonal to the collective mode $\tilde
\zeta_{\sbm{k}_L}$. Since the correlations between the fluctuations for
$|\bm{x}|\ll L_c$ and those outside the local volume of ${\cal O}(L_c^3)$ are negligibly small, in what follows, we neglect the fluctuations outside the local volume, which will disappear after taking the limit $L_c \to \infty$.

The commutation relation of $Q_\zeta^W$ with the long mode
$\zeta_{\sbm{k}_L}$ is given by
\begin{equation}
 \left[ i Q_\zeta^W(\bm{k}_L)\, ,\,\,\tilde{\zeta}_{\sbm{p}_L}  \right] = 
     - (2\pi L_c)^3 s_{-\sbm{k}_L} \hat {\cal W}(\bm{k}_L + \bm{p}_L)\,.
  \label{Eq:ComkL} 
\end{equation}
Equation \eqref{Eq:ComkL} states that the
generator $Q_\zeta^W(\bm{k}_L)$ 
shifts the collective soft mode $\tilde{\zeta}_{\sbm{k}_L}$
by $- s_{\sbm{k}_L} (2 \pi L_c)^3$. We will find that the factor 
$(2 \pi L_c)^3$, which blows up in the limit $L_c \to \infty$, is
cancelled out in the final expression of the consistency relation as it
should be. Whilst, the commutation relation with the short mode
$\zeta_{\sbm{p}_S}$ is given by
\begin{align}
 & \left[ i Q_\zeta^W(\bm{k}_L)\, ,\,\,\zeta_{\sbm{p}_S}  \right] \simeq
 s_{- \sbm{k}_L}\, \partial_{\sbm{p}_S} ( \bm{p}_S\, \zeta_{\sbm{p}_S+ \sbm{k}_L} ) \label{Eq:ComkS}\,,
\end{align}
where we approximated $\bm{k}_S + \bm{k}_L$ as $\bm{k}_S$.  

Repeating the argument around Eqs.~(\ref{Exp:psiz00})-(\ref{Def:ehzr})
except that the homogeneous mode $\bar{\zeta}$ is now replaced with a collective  
inhomogeneous soft modes $\tilde{\zeta}_{\sbm{k}_L}$, we expand the quantum state
$|\, \Psi \rangle$ in terms of the orthonormal basis 
$\{ |\, \tilde{\zeta}^c_{\sbm{p}_L} \rangle \}$, which are the
eigenstates of $\tilde{\zeta}_{\sbm{p}_L}$, as  
\begin{align}
 & |\Psi \rangle = \int d
 \tilde{\zeta}^c_{\sbm{p}_L} \,
  | \psi(\tilde{\zeta}^c_{\sbm{p}_L})|\,
  |\, \tilde{\zeta}^c_{\sbm{p}_L}\, \rangle
  |\, \Psi \rangle_{\tilde{\zeta}^c_{\sbm{p}_L}}  \label{Exp:psiz0}\,,
\end{align}
where 
we factorized the amplitude of the wave function,
$|\psi(\tilde{\zeta}^c_{\sbm{k}_L})|$, from $\langle
\tilde{\zeta}_{\sbm{k}_L}^c | \Psi \rangle$, as before. Using Eq.~(\ref{Eq:ComkL}), we obtain
\begin{align}
 & i Q_\zeta^W(\bm{k}_L) |\,
 \tilde{\zeta}^c_{-\sbm{k}_L}\, \rangle  = (2\pi L_c)^3  s_{- \sbm{k}_L} 
 \frac{\partial}{\partial
 \tilde{\zeta}^c_{-\sbm{k}_L}} |\, \tilde{\zeta}^c_{-\sbm{k}_L}\, \rangle\,.  \label{Eq:soft}
\end{align}
As we have already discussed, the dilatation invariance requires the
conditions (\ref{Cond:h0}) and (\ref{Cond:inh0}). In particular, the
second condition (\ref{Cond:inh0}) restricts how the inhomogeneous modes
respond to the insertion of the homogeneous mode $\bar{\zeta}$.  In the following, we will show that when this condition
can be extended to the soft mode with $\bm{k}_L \neq 0$, i.e., 
\begin{align}
 & i Q_\zeta^W(\bm{k}_L)  |\, \Psi \rangle_{\tilde{\zeta}^c_{-\sbm{k}_L}}
 = (2\pi L_c)^3  s_{  -\sbm{k}_L}  
  \frac{\partial} {\partial \tilde{\zeta}^c_{-\sbm{k}_L}}
   |\, \Psi \rangle_{\tilde{\zeta}^c_{-\sbm{k}_L}}  \label{Cond:hard}
\end{align}
is fulfilled, we can derive the consistency relation for $\zeta$ as
shown below.

In order to show the consistency relation by using the condition
(\ref{Cond:hard}), we evaluate 
\begin{align}
 & \langle \Psi\,|\, [i Q_\zeta^W(\bm{k}_L),\, \zeta_{\sbm{k}_{S1}}
 \cdots \zeta_{\sbm{k}_{Sn}} ] \,|\, \Psi \rangle 
\end{align}
in two ways: first by operating $i Q_\zeta^W(\bm{k}_L)$ on the quantum
state $|\, \Psi \rangle$ and second by considering the change of the
short wavelength modes $\zeta_{\sbm{k}_S}$ under the inhomogeneous dilatation,
expressed by the commutation relation with $i Q_\zeta^W(\bm{k}_L)$. When the condition
(\ref{Cond:hard}) is satisfied, using Eq.~(\ref{Eq:soft}), we obtain
\begin{align}
 i  Q_\zeta^W(\bm{k}_L) |\, \Psi \rangle &
 = (2\pi L_c)^3 s_{- \sbm{k}_L} 
  \int d \tilde{\zeta}^c_{-\sbm{k}_L} \, 
  |\psi(\tilde{\zeta}^c_{-\sbm{k}_L})|\,
 \frac{\partial}{\partial \tilde{\zeta}^c_{-\sbm{k}_L}}
  \left( |\, \tilde{\zeta}^c_{-\sbm{k}_L}\, \rangle |\,
   \Psi \rangle_{\tilde{\zeta}^c_{-\sbm{k}_L}}  \right) \cr
 & = - (2\pi L_c)^3 s_{- \sbm{k}_L}
  \int d \tilde{\zeta}^c_{-\sbm{k}_L}\, 
  \frac{\partial
 |\psi(\tilde{\zeta}^c_{-\sbm{k}_L})|}{\partial
 \tilde{\zeta}^c_{- \sbm{k}_L}}\, 
  |\,\tilde{\zeta}^c_{-\sbm{k}_L}\, \rangle 
   |\, \Psi \rangle_{\tilde{\zeta}^c_{-\sbm{k}_L}} \,,
\end{align}

When we neglect the non-linear contributions of the soft
modes\footnote{Here,  all the interaction vertexes 
connected to more than two soft modes are neglected.}, the wave
function of $\tilde{\zeta}^c_{\sbm{k}_L}$ is given by the Gaussian distribution function. Since the
square of $|\psi (\tilde{\zeta}^c_{\sbm{k}_L})|$ gives the probability
distribution, we can express the amplitude of the Gaussian wave function
as
\begin{align}
 & |\psi (\tilde{\zeta}^c_{-\sbm{k}_L})| \propto \exp \left( -
 \frac{\tilde{\zeta}^c_{\sbm{k}_L}
 \tilde{\zeta}^c_{-\sbm{k}_L} }{4 ( 2\pi L_c)^3 P_\zeta(k_L)} \right)\,, \label{Exp:psikL}
\end{align}
where $P_\zeta(k_L)$ denotes the power spectrum of $\zeta$. 
This is because the variance 
of $\tilde{\zeta}^c_{\sbm{k}_L}$ is given by 
\begin{align}
 & \langle \,|\,\tilde{\zeta}_{\sbm{k}_L}\,|^2 \rangle 
 = (2 \pi)^3 L_c^6 \int d^3 \bm{k} \hat{\cal W}(\bm{k}_L - \bm{k})
 \hat{\cal W}(-\bm{k}_L + \bm{k}) P_\zeta(k_L) \simeq (2 \pi L_c)^3
 P_\zeta(k_L) \,,
\end{align}
where the last equality is exact 
in the limit $L_c \to \infty$. Remember that $\tilde{\zeta}^{c}_{\sbm{k}_L}$ is not 
an independent variable from $\tilde{\zeta}^{c}_{-\sbm{k}_L}$, 
since $\tilde{\zeta}^{c}_{\sbm{k}_L}=\tilde{\zeta}^{c*}_{-\sbm{k}_L}$ holds from 
the reality of $\zeta$.

Using Eq.~(\ref{Exp:psikL}), we obtain
\begin{align}
 & \frac{\partial}{\partial \tilde{\zeta}^c_{-\sbm{k}_L}} 
  | \psi (\tilde{\zeta}^c_{-\sbm{k}_L}) | \SL -
 \frac{ \tilde{\zeta}^c_{\sbm{k}_L}}{2 ( 2\pi L_c)^3 P_\zeta(k_L)}\, | \psi
 (\tilde{\zeta}^c_{- \sbm{k}_L})|\,
 \,. \label{Eq:psikL}
\end{align}
Here and hereafter, we use $\SL$ to express that we approximate the wave function of the soft mode by the above Gaussian distribution function.
Replacing the eigenvalue $\tilde{\zeta}^c_{\sbm{k}_L}$ with the operator
$\tilde{\zeta}_{\sbm{k}_L}$, which commutes with the integral over $\tilde{\zeta}^c_{\sbm{k}_L}$, we obtain  
\begin{align}
 &  i  Q_\zeta^W(\bm{k}_L) |\, \Psi \rangle  \SL 
 \frac{s_{- \sbm{k}_L}}{2 P_\zeta(k_L)} \, \tilde \zeta_{\sbm{k}_L} |\, \Psi \label{Exp:opQ}
 \rangle \,.
\end{align}
Using this expression, we arrive at
\begin{align}
 & \langle \Psi \, | [i Q_\zeta^W(\bm{k}_L),\, \zeta_{\sbm{k}_{S 1}}
 \cdots  \zeta_{\sbm{k}_{S n}} ] |\, \Psi \rangle 
 \SL -
 \frac{ s_{- \sbm{k}_L}}{P_\zeta(k_L)} 
   \langle  \Psi \, | \tilde\zeta_{\sbm{k}_L} \zeta_{\sbm{k}_{S 1}}
 \cdots  \zeta_{\sbm{k}_{S n}} |\, \Psi \rangle \,.
\end{align}
Meanwhile, using Eq.~(\ref{Eq:ComkS}), we obtain  
\begin{align}
 & \langle \Psi \, | [i Q_\zeta^W(\bm{k}_L),\, \zeta_{\sbm{k}_{S 1}}
 \cdots  \zeta_{\sbm{k}_{S n}} ] |\, \Psi \rangle 
 =  s_{- \sbm{k}_L} \sum_{i=1}^n
 \partial_{\sbm{k}_{Si}} \bm{k}_{Si}  \langle \Psi \, | \zeta_{\sbm{k}_{S 1}}
 \cdots  \zeta_{\sbm{k}_{S n}} |\, \Psi \rangle\,.
\end{align}
Equating these two expressions derived from the two different ways, 
we have 
\begin{align}
 -  \langle  \Psi \, | \tilde\zeta_{\sbm{k}_L} \zeta_{\sbm{k}_{S 1}}
 \cdots  \zeta_{\sbm{k}_{S n}} |\, \Psi \rangle 
 \SL
  {P_\zeta(k_L)}  \sum_{i=1}^n
 \partial_{\sbm{k}_{Si}} \bm{k}_{Si}  \langle \Psi \, | \zeta_{\sbm{k}_{S 1}}
 \cdots  \zeta_{\sbm{k}_{S n}} |\, \Psi \rangle\,.
\label{averagedCR}
\end{align}
The ordinary form of the consistency relation (\ref{Exp:CRd}) is 
the one obtained by removing 
the delta function, which describes the momentum conservation,  
from the expression 
\begin{align}
 -  \langle  \Psi \, | \zeta_{\sbm{p}_L} \zeta_{\sbm{k}_{S 1}}
 \cdots  \zeta_{\sbm{k}_{S n}} |\, \Psi \rangle 
 \SL
  {P_\zeta(p_L)}  \sum_{i=1}^n
 \partial_{\sbm{k}_{Si}} \bm{k}_{Si}  \langle \Psi \, | \zeta_{\sbm{k}_{S 1}}
 \cdots  \zeta_{\sbm{k}_{S n}} |\, \Psi \rangle\,.
 \label{CRoriginal}
\end{align}
If we take the average of Eq.\eqref{CRoriginal} operating 
$L_c^3 \int d^3\bm{p}_L \hat {\cal W}(\bm{k}_L-\bm{p}_L)$, 
we recover Eq.~\eqref{averagedCR}. After we take the limit $L_c\to \infty$, which is automatically 
required when we take the soft limit $\bm{k}_L\to 0$, the averaging window 
becomes infinitesimally narrow. Therefore, we can conclude 
that Eq.~\eqref{averagedCR} is 
equivalent to the ordinary consistency relation. 

Under the Gaussian approximation of 
$|\, \psi (\tilde{\zeta}^c_{\sbm{k}_L})\,|$,  
the only assumption imposed to derive the consistency relation
(\ref{Exp:CRd}) is Eq.~(\ref{Cond:hard}), which states
that the influence of inhomogeneous dilatation is identical to shifting, by $+(2\pi L_c)^3 s_{\sbm{k}_L}$,
the collective soft mode $\tilde\zeta_{\sbm{k}_L}$ which
interacts with the hard modes. 
(The factor $(2\pi L_c)^3$ here is an artifact caused by discussing the Fourier space collective mode.)  
The consistency relation can be obtained, when the
condition (\ref{Cond:inh0}), which was required to preserve the dilatation invariance of the quantum
state, can be extended to the inhomogeneous soft mode with 
$\bm{k}_L \neq 0$. In this sense, the condition (\ref{Cond:hard}) can be
understood as a quantum version of the condition for Weinberg's adiabatic mode. The
condition (\ref{Cond:hard}) cannot be satisfied, in case the (linear)
soft mode $\zeta_{\sbm{k}_L}$ does not stop evolving in time in the
limit $\bm{k}_L \to 0$. In fact, the condition (\ref{Cond:hard}) states that performing the inhomogeneous dilatation transformation that 
induces the time-independent shift of $\zeta_{\sbm{k}_L}$, is equivalent
to shifting (the eigen value of) $\zeta_{\sbm{k}_L}$ 
at the evaluation time, which implies that the inserted
soft mode should be dominated by a constant contribution.

Now, we show that the condition (\ref{Cond:hard}) can be
indeed understood as the locality condition. For this purpose, 
we consider a set of eigenstates for the Fourier mode
$\zeta_{\sbm{k}_L}$ which satisfies
$\zeta_{\sbm{k}_L}|\,\zeta^c_L\,\rangle= \zeta^c_{\sbm{k}_L}|\,\zeta^c_L\,\rangle$ 
instead of the eigenstates for the collective mode
$\tilde{\zeta}_{\sbm{k}_L}$. Using the complete set of
$|\,\zeta^c_L\,\rangle$, we decompose the quantum state
$|\,\Psi\,\rangle$ as
\begin{equation}
  |\,\Psi\,\rangle=\int {\cal D}\zeta_{L} |\,\psi(\zeta^c_L\,)|\,
   |\,\zeta^c_L\,\rangle\,
    |\,\Psi\,\rangle_{\zeta^c_L}\,, 
\end{equation}
where ${\cal D}\zeta_{L}$ represents the functional integral over all soft modes. 
The extension of 
Eq.~(\ref{Cond:hard}) to the case of continuous modes will be 
\begin{align}
 & i Q_\zeta^W(\bm{k}_L)  |\, \Psi \rangle_{{\zeta}^c_{L}}
 =  s\hat W( -\bm{k}_L)   
  \frac{\delta} {\delta {\zeta}^c_{-\sbm{k}_L}}
   |\, \Psi \rangle_{{\zeta}^c_{L}}\,.  \label{Cond:hard2}
\end{align}
Using this relation, we obtain
\begin{align}
i Q_\zeta^W(\bm{x})  |\,\Psi\,\rangle_{\zeta^c_L}\, 
 &  =s 
 \int \frac{d^3 \bm{k}_L}{(2\pi)^3} e^{-i \sbm{k}_L \cdot \sbm{x}}\,
 \hat W( \bm{k}_L)   \frac{\delta}{\delta \zeta_{\sbm{k}_L}}  |\,\Psi\,\rangle_{\zeta^c_L}\, = s \frac{\delta}{\delta \zeta_L(\bm{x})}  |\,\Psi\,\rangle_{\zeta^c_L}\,\,, 
  \label{Cond:hardlc}
\end{align}
where 
\begin{equation}
 \zeta_L(\bm{x})\equiv 
    \int\frac{d^3\bm{k}_L}{(2\pi)^3} 
    \hat W( \bm{k}_L)  e^{i\sbm{k}_L \cdot \sbm{x}}\zeta_{\sbm{k}_L}
 \,
 \end{equation}
is the coarse grained field corresponding to the degrees of freedom of soft modes. 
Let us imagine a set of the separate universes whose sizes are of ${\cal O}(L)$. 
The operator $Q_\zeta^W(\bm{x})$ induces the dilatation
only within the separate universe centered at $\bm{x}$. 
The condition 
(\ref{Cond:hardlc}) states that the impact of the soft mode on the quantum state of short modes 
is limited only to the influence which is equivalent to the 
inhomogeneous dilatation in the separate universe.

Notice that since (the amplitude of) the wave function is not flat in the direction of the
inhomogeneous mode of $\zeta$, i.e., 
$
\partial |\psi(\tilde{\zeta}_{\sbm{k}_L})|/\partial \tilde{\zeta}_{\sbm{k}_L} \neq 0\,,
$
as seen in Eq.~(\ref{Eq:psikL}), the quantum state $|\, \Psi \rangle$
does not remain invariant under the inhomogeneous dilatation 
$Q_\zeta^W(\bm{k}_L)$, i.e., 
\begin{align}
 & Q_\zeta^W(\bm{k}_L) |\, \Psi \rangle  \neq 0\,. \label{Changezeta}
\end{align} 
As shown in Eq.~(\ref{Exp:opQ}), operating the generator of the
inhomogeneous dilatation inserts the soft mode
$\tilde{\zeta}_{\sbm{k}_L}$ with $\bm{k}_L \neq 0$, which changes the quantum
state, even if the state is invariant under the homogeneous dilatation. As will be
discussed in Sec.~\ref{Sec:IR}, because of that, choosing a quantum
state which is invariant under the dilatation is not enough to guarantee
the IR regularity. Let us emphasize that there is no spontaneous
symmetry breaking for the large gauge transformations: the symmetry under the 
dilatation, generated by $Q_\zeta$, is preserved also after the quantization, 
while the
inhomogeneous dilatation, generated by $Q_\zeta^W$, is not a symmetry of
the classical action.

Although we assumed that the wave function of the soft mode,
$|\psi(\tilde{\zeta}^c_{\sbm{k}_L})|$, is given by the Gaussian
distribution, the non-linear interactions of the hard modes are fully
kept. Therefore, the consistency relation thus derived can  
apply to a much more general setup compared to the original one by Maldacena in
Ref.~\cite{Maldacena02}. For instance, the wavelengths of the hard modes
can be arbitrary as far as $k_S L \simeq k_S/k_L \gg 1$, i.e., they are not
necessarily in super Hubble scales. In addition, obviously the same
argument as above can be applied, even if the time coordinates of
the short modes $\zeta_{\sbm{k}_{Si}}(t_i)$ with $i=1, \cdots, n$ are different
among them.

Because of the Gaussian approximation for $\psi(\tilde{\zeta}^c_{\sbm{k}_L})$,
the consistency relation derived here do not contain non-linear
interactions of the soft modes. Notice that since there is no
approximation for the soft modes in the locality condition
(\ref{Cond:hard}) or \eqref{Cond:hardlc}, when we include the non-linear contributions of the
soft modes, the same condition leads to the consistency
relation\footnote{When we include the non-linear contributions of the
soft modes, the dilatation also changes the argument of $\zeta$ and the transformation of $\zeta_{\sbm{k}_L}$ under the
dilatation is not the simple shift. In order not to change the argument, 
we need to introduce the window function in a
physical distance such as the geodesic distance.}. However, the non-linear interactions of the soft modes in general
yield the IR divergences through their radiative corrections. Therefore, this
extension requires a more careful consideration. This issue will be discussed in Sec.~\ref{Sec:IR}.

\section{Massive particles with arbitrary spins}  \label{Sec:spin}
In this section, we consider an influence of heavy fields with
arbitrary spins, which interact with the inflaton directly or 
indirectly through the gravitational interaction. The argument in this
section is a generalization of the one in Ref.~\cite{TY_conservation},
which showed the consistency relation for a heavy scalar field and the
conservation of $\zeta$, taking into account radiative corrections of
the heavy scalar field. In the previous section, we derived the
condition for the consistency relation, using the generator(s) of
the inhomogeneous dilatation. This argument can be
extended to the cases with the radiative corrections of massive non-zero
spin fields straightforwardly. In this section and in Appendix B, 
to keep generality, we
consider a $(\dd +1)$-dimensional spacetime.

\subsection{Setup of the problem}
In this section, we consider a heavy field whose mass is $M_S$ and
spin is $S \geq 0$, including higher spin fields. We only consider the
mass range where there is no instability~\cite{NJ15, Deser:2001us},
e.g., $M_2 \geq 2 H$ for a spin 2 field. For our purpose, we do not need to specify the detail
of the interaction for the interacting system with the inflaton and the
massive fields. We simply express the action as
\begin{align}
 & S[\delta g,\, \bch]= S_{ad}[\delta g] + S_\chi[\delta g,\, \bch] \,,
\end{align}
with
\begin{align}
 & S_\chi[\delta g,\, \bch] \equiv \sum_\alpha \int d t\, d^\dd \bm{x}\, a^\dd\, e^{\dd \zeta(x)}
 f^{\{i_\alpha\}}(\phi,\, \delta g) O_{\{i_\alpha\}}(x)\,,  \label{Exp:actionspin}
\end{align}
where $\bch$ and $O_{{i_\alpha}}$ denote a set of heavy fields $\chi^I$ with $I=1,\,2,\, \cdots$ 
and a composite operator of $\bch$, respectively, and $\delta g$ denotes
the set of the metric perturbations, $N$, $N_i$, $\zeta$, and
$\gamma_{ij}$. Here, the action $S_{ad}[\delta g]$ only includes the metric
perturbations in the unitary gauge defined by the condition 
$\delta \phi=0$ and it is identical to the action in single
field models of inflation.

The heavy fields and the inflaton $\phi$ also can interact
 directly. 
 For the present purpose, 
 we do not need to specify the composite operators
 $O_{{i_\alpha}}$. 
 We only need to specify their scaling
 dimensions $\Delta_\alpha$, i.e., they transform as
\begin{align}
 & O^s_{\{i_\alpha\}}(t,\, \bm{x}_s) = e^{-\Delta_\alpha s}
 O_{\{i_\alpha\}} (t,\, \bm{x})  \label{TransO}
\end{align}
under the dilatation transformation $\bm{x} \to \bm{x}_s = e^s \bm{x}$,
where ${i_\alpha}$ denotes tensor indices. While the heavy fields can be
a fermion with a half integer spin, we assume that $\delta g$ interacts
with $\bch$ only through the composite
operators which transform as tensors (with an integer spin) under coordinate
transformations. A composite operator with $n$ tensor (lower) indices has a
scaling dimension $n$, e.g., $\Delta=0$ for a scalar composite operator
and $\Delta=1$ for a vector one.

\subsection{Soft theorem for heavy fields with non-zero spins}
In this subsection, we derive the consistency relation or the soft
theorem in the presence of the heavy fields, extending the discussion in the previous
section. In the previous section, expanding the quantum state 
$|\, \Psi \rangle$ as in Eq.~(\ref{Exp:psiz00}), we derived the conditions
(\ref{Cond:h0}) and (\ref{Cond:inh0}) by requesting the invariance of
the quantum state under the dilatation. Here, repeating the same
argument except that now the quantum state $|\, \Psi \rangle$ also
includes the heavy fields in addition to the inhomogeneous modes of $\zeta$, we obtain
the same conditions as Eqs.~(\ref{Cond:h0}) and (\ref{Cond:inh0}) from
the dilatation invariance of $|\, \Psi \rangle$. Since 
$\ehzr$ also includes the degrees of freedom for the heavy fields, the
condition (\ref{Cond:inh0}) restricts how both of the inhomogeneous
modes of $\zeta$ and the heavy fields should respond to the dilatation
transformation. Similarly, when the condition (\ref{Cond:inh0}) can be
extrapolated to the inhomogeneous dilatation with the fall-off at the
spatial infinity by replacing the homogeneous mode $\bar{\zeta}^c$ in
(\ref{Cond:inh0}) with the soft mode $\tilde{\zeta}_{\sbm{k}_L}$, 
i.e., when the locality condition (\ref{Cond:hard}) is fulfilled also in the
interacting system with the inflaton and the heavy fields, we can derive
the soft theorem, which describe the influence of the soft
mode.

Recall that the consistency relation can be derived by evaluating the
change of the quantum state $|\, \Psi \rangle$ and the change of the
operators for the heavy fields. In order to derive the soft theorem for
the heavy fields, we consider
\begin{align}
 & \langle \Psi \,| [ iQ_\zeta^W(\bm{k}_L),\, O_{\{i_{\alpha_1}\}
 \sbm{k}_{S1}}(t_1)  \cdots  O_{\{i_{\alpha_n}\} \sbm{k}_{Sn}}(t_n)] |\,
 \Psi \rangle\,, 
\end{align}
where $O_{\{i_{\alpha}\} \sbm{k}_{S}}(t)$ denotes the Fourier mode
of $O_{\{i_{\alpha}\}}(x)$ with $k_L/k_S \ll 1$. Repeating the same argument and taking the limit $L_c \to \infty$, we find that the condition
(\ref{Cond:hard}) implies Eq.~(\ref{Exp:opQ}). Then, using
Eq.~(\ref{Exp:opQ}), we can compute the change of the quantum state
under the inhomogeneous dilatation as
\begin{align}
 & \langle \Psi \, | [i Q_\zeta^W(\bm{k}_L),\,  O_{\{i_{\alpha_1}\} \sbm{k}_{S1}}(t_1)  \cdots  O_{\{i_{\alpha_n}\} \sbm{k}_{Sn}}(t_n) ] |\, \Psi \rangle \cr
 & \quad \SL - \frac{s_{- \sbm{k}_L}}{P_\zeta(k_L)} \langle  \Psi \, |
 \zeta_{\sbm{k}_L} O_{\{i_{\alpha_1}\} \sbm{k}_{S1}}(t_1)  \cdots
 O_{\{i_{\alpha_n}\} \sbm{k}_{Sn}}(t_n) |\, \Psi \rangle \,. 
\end{align}
We equate this expression with the one obtained by computing the change of
the composite operators $O_{\{i_{\alpha}\} \sbm{k}}$ under the
(inhomogeneous) dilatation. Since $O_{\{i_{\alpha}\} \sbm{k}}$
transforms as in Eq.~(\ref{TransO}), we obtain
\begin{align}
 & \left[ i Q_\zeta^W(\bm{k}_L),\, O_{\{i_{\alpha_1}\} \sbm{k}_S} \right]
 = s_{- \sbm{k}_L} \left( \frac{\partial}{\partial \bm{k}_S} \bm{k}_S - \Delta_\alpha
 \right)  O_{\{i_{\alpha_1}\} \sbm{k}_S} \,.  
\end{align} 
Likewise in the discussion for the single field case, whether the
dilatation parameter is homogeneous or inhomogeneous does not affect the transformation of the short
modes. Using this expression, we obtain
\begin{align}
 & \langle \Psi \, | [i Q_\zeta^W(\bm{k}_L),\,  O_{\{i_{\alpha_1}\}
 \sbm{k}_{S1}}(t_1)  \cdots  O_{\{i_{\alpha_n}\} \sbm{k}_{Sn}}(t_n) ]
 |\, \Psi \rangle \cr 
 & \quad \SL s_{- \sbm{k}_L}  \sum_{i=1}^n
 \left( \partial_{\sbm{k}_{Si}} \bm{k}_{Si} - \Delta_{\alpha_i} \right)
 \langle \Psi \, |  O_{\{i_{\alpha_1}\} \sbm{k}_{S1}}(t_1)  \cdots  O_{\{i_{\alpha_n}\} \sbm{k}_{Sn}}(t_n) |\, \Psi \rangle \,, 
\end{align}
Equating these two expressions and sending $L_c$ to the infinity, we obtain the consistency relation for
the heavy fields as
\begin{align}
 & \lim_{k_L \to 0} \frac{\langle \Psi | \zeta_{\sbm{k}_L}  O_{\{i_{\alpha_1}\} \sbm{k}_{S1}}(t_1)  \cdots  O_{\{i_{\alpha_n}\} \sbm{k}_{Sn}}(t_n) |
 \Psi \rangle' }{P_\zeta(k_L)} \cr
 & \, \SL  -  \left(  \sum_{i=2}^n  \bm{k}_{Si} \cdot \frac{\partial}{\partial
 \bm{k}_{Si}} + d(n-1) - \Delta  \right)  \langle \Psi | O_{\{i_{\alpha_1}\} \sbm{k}_{S1}}(t_1)  \cdots  O_{\{i_{\alpha_n}\}
 \sbm{k}_{Sn}}(t_n)  |
 \Psi \rangle'
\,, \label{Exp:CRdgO}
\end{align}
where we defined $\Delta \equiv \sum_{i=1}^n \Delta_{\alpha_i}$. We put
a prime to denote the correlation functions without the
multiplicative factor $(2 \pi)^\dd$ and the delta function which expresses
the momentum conservation.

\subsection{Effective action}
In order to show that the radiative corrections of the heavy fields do
not induce any time evolution of $\zeta$ at large scales when the
condition (\ref{Cond:hard}) is satisfied, we compute the effective
action for $\zeta$ by integrating out the heavy fields $\bch$ in the closed time path (or the in-in) formalism. In
particular, the contributions of the heavy fields $\bch$ are described
by the Feynman and Vernon's influence functional~\cite{FV, FH}. In this subsection, we briefly
summarize the way to calculate the influence functional and the effective
action. We will see that now the argument to show the conservation proceeds
almost in parallel to the one for the heavy scalar field, discussed in
Ref.~\cite{TY_conservation}.

\subsubsection{Influence functional}
Performing the path integral along the closed time path, the $n$-point
function of the curvature perturbation $\zeta$ is given by 
\begin{align}
 & \langle \Psi\,| T\zeta(x_1) \cdots \zeta(x_n) |\, \Psi \rangle\cr
 & = \frac{\int D \delta g^{dy}_+ \int D\bch_+ \int D \delta g^{dy}_- \int D\bch_-\,
 \zeta_+(x_1) \cdots \zeta_+(x_n)\, e^{i S[\delta g_+,\, \sbch_+]-
 i S[\delta g_-,\, \sbch_-]}}{\int D \delta g^{dy}_+ \int D\bch_+ \int D
 \delta g^{dy}_- \int
 D\bch_-\, e^{i S[\delta g_+,\, \sbch_+]-
 i S[\delta g_-,\, \sbch_-]}}\,,
\end{align}
where we double the fields: $\delta g_+$ and
$\bch_+$ denote the fields defined along the path from the past infinity
to the time $t$ and $\delta g_-$ and $\bch_-$ denote the fields
integrated from the time $t$ to the past infinity. Since $N$ and 
$N_i$ are the Lagrange multiplies, which are eliminated by solving the
constraint equations, we perform the path integral only regarding the dynamical
degrees of freedom $\delta g^{dy} \equiv (\zeta,\, \gamma_{ij})$ and 
$\bch$. An insertion of $\delta g_+(x)$ into the path integral in the
numerator as above gives a correlation function in the time ordering,
expressed by $T$, while an insertion of $\delta g_-(x)$ gives a
correlator in the anti-time ordering, expressed by $\bar{T}$.

Separating the part which describes the radiative corrections of the
heavy fields as
\begin{align}
 &i \Seff[\delta g_+,\, \delta g_-] \equiv  \ln \left[ \int D \chi_+ \int D \chi_- \,
 e^{i S[\delta g_+,\, \chi_+]- i S[\delta g_-,\, \chi_-]}\right]\,, \label{Def:Seff}
\end{align}
we can express the $n$-point function for $\zeta$ superficially as if
there are only the metric perturbations and the inflaton as
\begin{align}
 &  \langle \Psi\,| T\zeta(x_1) \cdots \zeta(x_n) |\, \Psi \rangle
 = \frac{\int D \delta g^{dy}_+  \int D \delta g^{dy}_- \, \zeta_+(x_1) \cdots
 \zeta_+(x_n)\, e^{i \Seff[\delta g_+,\,
 \delta g_-]}}{\int D \delta g^{dy}_+ \int D \delta g^{dy}_-\,e^{i
 \Seff[\delta g_+,\,\delta g_-]}} \,.   
\end{align}
The effective action is recast into
\begin{align}
 & \Seff[\delta g_+,\, \delta g_-] = S_{ad}[\delta g_+] -
 S_{ad}[\delta g_-] + \Seff'[\delta g_+,\, \delta g_-]\,, \label{Exp:Seff}
\end{align}
where $\Seff'$ is the so-called influence functional, given by 
\begin{align}
 &i   \Seff'[\delta g_+,\, \delta g_-] 
 \equiv \ln \left[ \int D \bch_+ \int D \bch_- \,e^{i S_\chi[\delta g_+,\,
 \sbch_+]- i S_\chi[\delta g_-,\, \sbch_-]}\right]\,,  \label{Exp:Seffd}
\end{align}
where we factorized $S_{ad}[\delta g_\pm]$ which commutes with the path integral
over $\bch_\pm$. Here, we only consider the correlation functions for
$\zeta$, but the effective action $\Seff[\delta g_+,\, \delta g_-]$ describes the
evolution of both $\zeta$ and $\gamma_{ij}$ affected by the quantum
fluctuations of the heavy fields $\bch$. For our later use, we introduce
the correlation functions of $\bch$ computed in the absence of the
metric perturbations as 
\begin{align}
 & \langle {\cal O}[\bch_+,\, \bch_-] \rangle_{\pm} \equiv \frac{\int D \bch_+
 \int D \bch_- \,{\cal O}[\bch_+,\, \bch_-] e^{i S_\chi[0,\, \sbch_+]- i
 S_\chi[0,\,\sbch_-]}}{\int D \bch_+ \int D \bch_- \,e^{i S_\chi[0,\,\sbch_+]- i
 S_\chi[0,\,\sbch_-]}} \,.
\end{align}

Expanding $\Seff'$ in terms of the metric perturbations $\delta g=
(\delta N,\, N_i,\, \zeta, \, \gamma_{ij})$, we obtain 
\begin{align}
 & i \Seff'[\delta g_+,\, \delta g_-] \equiv \sum_{n=0}^\infty i
 \Seffd{n}[\delta g_+,\, \delta g_-] \,,
\end{align}
where $\Seffd{n}$ denotes the terms which include $n$
$\delta g_\alpha$s, given by
\begin{align}
 i \Seffd{n}[\delta g_+,\, \delta g_-] 
 &= \frac{1}{n!}
 \sum_{a_1 = \pm} \cdots \sum_{a_n = \pm}
 \int d^{\dd +1} x_1 \cdots \int d^{\dd +1} x_n \cr
 & \qquad \times \delta g_{a_1}(x_1) \cdots
 \delta g_{a_n}(x_n)\,W^{(n)}_{\delta g_{a_1}
 \cdots \delta g_{a_n}}(x_1,\, \cdots,\, x_n) \,, \label{Expn:Seffd}
\end{align} 
with the non-local interaction vertices induced by the heavy fields:
\begin{align}
 & W^{(n)}_{\delta g_{a_1} \cdots \delta g_{a_n}}(x_1,\, \cdots,\, x_n) \equiv
 \frac{\delta^n i \Seff'[\delta g_+,\, \delta g_-]}{\delta g_{a_1}(x_1)
 \cdots \delta  g_{a_n}(x_n)}
 \bigg|_{\delta g_{\pm}=0} \,.  \label{Def:tWn}
\end{align}
In Eq.~(\ref{Expn:Seffd}), each $\delta g_{a_m}$ with 
$m=1,\cdots,\, n$ should add up all the metric perturbations $\delta N_{a_m}$, 
$N_{i, a_m}$, $\zeta_{a_m}$, $\gamma_{ij\, a_m}$. Here and
hereafter, for notational brevity, we omit the summation symbol over
$\delta g$ unless necessary. Inserting Eq.~(\ref{Exp:Seffd}) into
Eq.~(\ref{Def:tWn}), we can express the non-local interaction vertices
by using the correlators for $\bch$. These expressions are summarized in
Appendix ~\ref{Sec:IFcomp}. Once we expand the effective action
as in Eq.~(\ref{Expn:Seffd}), the shift symmetry is lost at each order
in perturbation about $\delta g$. In the following, we will show that we
can rewrite the effective action in such a way that the shift symmetry is
manifestly preserved by using the consistency relation.

\subsubsection{Soft theorem and effective action}
In order to show the conservation of $\zeta$ in the presence of the
radiative corrections of the heavy fields, here we rewrite the consistency
relation (\ref{Exp:CRdgO}). As was mentioned in the previous section,
the time coordinates of the hard modes can be different among different
composite operators. Taking an appropriate ordering of the composite
operators, i.e., we can put the index $a_i= \pm$ on each composite operator 
$O_{\{i_{\alpha_i}\} \sbm{k}_{Si}}$ in the consistency relation
(\ref{Exp:CRdgO}). In the following, we use the prescription introduced
in the previous subsection (see also Appendix 
\ref{Sec:IFcomp}). In particular, all the correlation functions should be
understood as being computed in the path ordering of the closed time
path with the distinction of $\pm$.

Employing the Gaussian approximation for the soft mode of $\zeta$ 
again, we can compute the
correlation function in the first line of Eq.~(\ref{Exp:CRdgO}) as
\begin{align}
 & \langle \Psi | \zeta_{\sbm{k}_L}  O^{a_1}_{\{i_{\alpha_1}\}
 \sbm{k}_{S1}} (t_1) \cdots  O^{a_n}_{\{i_{\alpha_n}\} \sbm{k}_{Sn}}(t_n) | \Psi
 \rangle \cr
 & \, \SL - i\,\, \langle \Psi,\, 0_{\zeta} \,|  S_\chi^{int\, -}\,
 \zeta_{\sbm{k}_L}\,  O^{a_1}_{\{i_{\alpha_1}\} 
 \sbm{k}_{S1}}(t_1)  \cdots  O^{a_n}_{\{i_{\alpha_n}\} \sbm{k}_{Sn}}(t_n) 
 |\, \Psi,\, 0_{\zeta}  \rangle  \cr
 & \qquad + i\,\, \langle \Psi,\, 0_{\zeta} \,|
 \zeta_{\sbm{k}_L}\,  O^{a_1}_{\{i_{\alpha_1}\} 
 \sbm{k}_{S1}}(t_1)  \cdots  O^{a_n}_{\{i_{\alpha_n}\}
 \sbm{k}_{Sn}}(t_n) \,  S_\chi^{int\, +} |\, \Psi,\, 0_{\zeta}  \rangle
\,,
\end{align}
where $|\, \Psi,\, 0_{\zeta}  \rangle$ denotes the quantum state 
$|\, \Psi \rangle$
with $\zeta$ being in non-interacting vacuum. Here, $S_\chi^{int}$ denotes a set of the
interaction vertexes which include only one $\zeta$ without derivative and
the massive fields $\bch$ and is given by
\begin{align}
 & S_\chi^{int} = \int dt\, d^\dd \bm{x}\, \zeta(x) \frac{\delta
 S_\chi}{\delta \zeta(x)} \bigg|_{\zeta=0} = \int dt\,\int \frac{d^\dd
 \bm{k}}{(2\pi)^\dd}\, \zeta_{\sbm{k}}\, \frac{\delta S_\chi}{\delta
 \zeta}\bigg|_{\zeta=0} \hspace{-5pt} (t,\,-\bm{k}) 
\end{align}
with 
\begin{align}
 &  \frac{\delta S_\chi}{\delta \zeta}\bigg|_{\zeta=0} \hspace{-5pt}
 (t,\,\bm{k}) \equiv \int d^\dd \bm{x}\, e^{- i \sbm{k} \cdot \sbm{x}}  
  \frac{\delta S_\chi}{\delta \zeta(x)}\Big|_{\zeta=0} \,. 
\end{align}
We express the interaction action $S_\chi^{int}$ with the heavy fields
on the paths $\pm$ as $S_\chi^{int\, \pm}$, respectively. 
Factoring out the power spectrum of $\zeta$, we obtain
\begin{align}
 & \frac{\langle \Psi | \zeta_{\sbm{k}_L}  O^{a_1}_{\{i_{\alpha_1}\}
 \sbm{k}_{S1}}(t_1)  \cdots  O^{a_n}_{\{i_{\alpha_n}\}
 \sbm{k}_{Sn}}(t_n) | \Psi \rangle }{P_\zeta(k_L)} \cr
 &\, \SL  - i \int dt  \left\langle \Psi,\, 0_{\zeta} \Biggl|  \frac{\delta
 S^-_\chi}{\delta \zeta}\bigg|_{\zeta=0} \hspace{-5pt} 
 (t,\, \bm{k}_L) \, O^{a_1}_{\{i_{\alpha_1}\}
 \sbm{k}_{S1}}(t_1)  \cdots  O^{a_n}_{\{i_{\alpha_n}\}
 \sbm{k}_{Sn}}(t_n) \Biggr|\, \Psi,\, 0_{\zeta}  \right\rangle \cr
  & \qquad + i \int dt   \left\langle \Psi,\, 0_{\zeta} \Biggl| O^{a_1}_{\{i_{\alpha_1}\}
 \sbm{k}_{S1}}(t_1)  \cdots  O^{a_n}_{\{i_{\alpha_n}\}
 \sbm{k}_{Sn}}(t_n)\,  \frac{\delta
 S^+_\chi}{\delta \zeta}\bigg|_{\zeta=0} \hspace{-5pt} 
 (t,\, \bm{k}_L)  \Biggr|\, \Psi,\, 0_{\zeta}  \right\rangle \,, \label{tempp2}
\end{align}
where the power spectrum of $\zeta$ is canceled between the numerator
and the denominator. In the first line of Eq.~(\ref{tempp2}), we omitted the time coordinate of the soft mode
 $\zeta_{\sbm{k}_L}$, since it should be constant in time to
 satisfy the condition (\ref{Cond:hard}). (If one wants to specify the
 time coordinate of $\zeta_{\sbm{k}_L}$, we can place it along
 the closed time path at the end of inflation where
$\zeta_{\sbm{k}_L}^+=\zeta_{\sbm{k}_L}^-$.) Expressing the
composite operators in the position space, we obtain 
\begin{align}
 &  \sum_{i=1}^n \{ \partial_{\sbm{x}_i} \bm{x}_i -(d- \Delta_i) \}
 \left\langle \Psi,\, 0_{\zeta} \Biggl| O^{a_1}_{\{i_{\alpha_1}\}}(x_1)
 \cdots  O^{a_n}_{\{i_{\alpha_n}\}}(x_n) \Biggr|\, \Psi,\, 0_{\zeta}
 \right\rangle \cr  &\, \SL  - i \int dt  \left\langle \Psi,\, 0_{\zeta} \Biggl|  \frac{\delta
 S^-_\chi}{\delta \zeta}\bigg|_{\zeta=0} \hspace{-5pt} 
 (t,\, \bm{k}_L)\,  O^{a_1}_{\{i_{\alpha_1}\}}(x_1)  \cdots  {\cal
 O}^{a_n}_{\{i_{\alpha_n}\}}(x_n)  \Biggr|\, \Psi,\, 0_{\zeta}  \right\rangle \cr
  & \qquad + i \int dt   \left\langle \Psi,\, 0_{\zeta} \Biggl| {\cal
 O}^{a_1}_{\{i_{\alpha_1}\}}(x_1)  \cdots  O^{a_n}_{\{i_{\alpha_n}\}}(x_n)  \frac{\delta
 S^+_\chi}{\delta \zeta}\bigg|_{\zeta=0} \hspace{-5pt} 
 (t,\, \bm{k}_L)  \Biggr|\, \Psi,\, 0_{\zeta}  \right\rangle \,. \label{Exp:WT}
\end{align}
Since we neglected the higher order contributions of the soft modes using the
approximation of $\SL$, we replaced the quantum state $|\, \Psi \rangle$ 
in the first line with $|\,  \Psi,\, 0_{\zeta}\rangle$.

\subsection{Conservation of $\zeta$}
For the purpose of showing the conservation of $\zeta$, let us introduce
\begin{align}
 & \delta \hat{g}(x)= \{ N(x),\, e^{-\zeta(x)} N_i(x),\, \dot{\zeta}(x),\, e^{-\zeta(x)}
 \partial_{\sbm{x}} \zeta(x),\, e^{-\zeta(x)} \partial_{\sbm{x}},\, e^{-
 2 \zeta(x)} \gamma_{ij}(x) \}\,.
\end{align}
Since the metric perturbations transform as
\begin{align}
 & N_s(t,\, \bm{x}_s) = N(t,\, \bm{x})\,, \qquad e^{s} N_{i,\,s}(t,\,
 \bm{x}_s) = N_i(t,\, \bm{x})\,, \label{Exp:transN} \\
 &\zeta_s(t,\, \bm{x}_s) + s = \zeta(t,\, \bm{x})\,,  \qquad e^{2s}
 \gamma_{ij\, s}(t,\, \bm{x}_s) = \gamma_{ij}(t,, \bm{x}) \label{Exp:transzeta}
\end{align}
under the dilatation transformation, we can easily see that 
$\delta \hat{g}(x)$ transform as a scalar under the
dilatation.

Taking into account that the composite operator $O_{\{i_\alpha\}}$
transforms as given in Eq.~(\ref{TransO}), we also can construct a
scalar operator for $O_{\{i_\alpha\}}$ as
$e^{-\Delta_\alpha \zeta(x)} O_{\{i_\alpha\}}(x)$. Here, introducing
\begin{align}
 &  \hat{f}^{\{i_\alpha\}} (\phi,\, \delta \hat{g}) \equiv
 e^{\Delta_\alpha \zeta}  f^{\{i_\alpha\}}(\phi,\, \delta g)\,
\end{align}
for each $f^{\{i_\alpha\}}$ in the action (\ref{Exp:actionspin}), we
rewrite the action for the heavy fields as 
\begin{align}
 & S_\chi[\delta g,\, \bch] \equiv \sum_a \int d t\, d^\dd \bm{x} \,a^\dd e^{\dd  \zeta(x)}
 \hat{f}^{\{i_\alpha\}}(\phi,\, \delta \hat{g}) e^{-\Delta_\alpha
 \zeta(x)} O_{\{i_\alpha\}}(x)\,. \label{Exp:actionspin2}
\end{align}
Since $\hat{f}^{\{i_a\}}$ transforms as a scalar under the dilatation, the
metric perturbations included in $\hat{f}^{\{i_a\}}$ can be expressed
only in terms of $\delta \hat{g}(x)$, which also transform as a scalar.

Taking the first and the second variations of $S_\chi$ with respect to $\zeta$, we obtain 
\begin{align}
 & \frac{\delta S_\chi}{\delta \zeta(x)} \bigg|_{\delta g=0} =
 \sum_{\alpha} a^\dd(t) (d- \Delta_\alpha) f^{\{i_\alpha\}}
 O_{\{i_\alpha\}}(x) \,, \\
 &  \frac{\delta^2 S_\chi}{\delta \zeta^2(x)} \bigg|_{\delta g=0} =
 \sum_{\alpha} a^\dd(t) (d- \Delta_\alpha)^2 f^{\{i_\alpha\}}
 O_{\{i_\alpha\}}(x)\,.
\end{align}
Since we expressed the action as in Eq.~(\ref{Exp:actionspin2}), the
terms in which the derivative operates on $\hat{f}^{\{i_\alpha\}}$
vanish. 
Variations with respect to
the other metric perturbations,  
$\delta N,\, e^{- \zeta} N_i,\, e^{- 2\zeta}\gamma_{ij}$, which are in
the combination of $\delta \hat{g}$, yield
\begin{align}
 & \frac{\delta S_\chi}{\delta \hat{g}(x)} \bigg|_{\delta g=0} =
 \sum_{\alpha} a^\dd(t)  \frac{ \partial \hat{f}^{\{i_\alpha\}}}{\partial
 \hat{g}(x)} \bigg|_{\delta g=0} O_{\{i_\alpha\}}(x) \,, \\
 &  \frac{\delta^2 S_\chi}{\delta \hat{g}(x) \delta \zeta(x)} \bigg|_{\delta g=0} =
 \sum_{\alpha} a^\dd(t) (d- \Delta_\alpha) \frac{ \partial \hat{f}^{\{i_\alpha\}}}{\partial
 \hat{g}(x)} \bigg|_{\delta g=0} O_{\{i_\alpha\}}(x) \,,
\end{align}
and so on. Multiplying 
$a^\dd(t) \partial \hat{f}^{\{i_\alpha\}}/\partial g(x)|_{\delta g =0}$
on Eq.~(\ref{Exp:WT}) with $n=1$ and all the remaining metric perturbations set
to 0 and taking summation over $\alpha$, we obtain 
\begin{align}
 & \partial_{\sbm{x}} \left\{   \bm{x}\, W^{(1)}_{\delta g_{\pm}}(x)
 \right\} \SL \int dt_y \int d^\dd \bm{y} e^{- i \sbm{k}_L \cdot \sbm{y}}
 \left\{ W^{(2)}_{\delta g_{\pm} \zeta_{+}}(x,\, y) + W^{(2)}_{\delta
 g_{\pm} \zeta_{-}}(x,\, y) \right\}  \label{Exp:WT0}
\end{align}
for $\delta g= \zeta,\, \delta N,\, N_i,\, \gamma_{ij}$. Notice that the
derivative with respect to $e^{- \zeta} N_i$ and the one with respect to
$N_i$ give the same answer after setting $\delta g$ to 0. The same story
also follows for $\gamma_{ij}$. Here, we
expressed $\delta S_\chi/\delta \zeta|_{\zeta=0}(t,\, \bm{k}_L)$ in
Eq.~(\ref{Exp:WT}) in the position space. In deriving
Eq.~(\ref{Exp:WT}), we used the explicit
forms of $W^{(2)}_{\delta g_{a_1} \delta \tilde{g}_{a_2}}(x_1,\, x_2)$,
summarized in Appendix \ref{Sec:IFcomp}. Multiplying $\delta g_{\pm}$
on Eq.~(\ref{Exp:WT0}) and integrating over $x^\mu$, we finally obtain 
the key formula to show the presence of the constant solution as
\begin{align}
 & \int d^{\dd +1} x\, \{\bm{x} \cdot \partial_{\sbm{x}} \delta g_{\pm} (x) \}
 W^{(1)}_{\delta g_{\pm}} (x) \cr
 & \qquad  + \int d^{\dd +1} x \int d^{\dd +1} y\,e^{- i \sbm{k}_L \cdot \sbm{y}} \delta g_{\pm}(x)
 \left\{ W^{(2)}_{\delta g_{\pm} \zeta_{+}}(x,\, y) + W^{(2)}_{\delta
 g_{\pm} \zeta_{-}}(x,\, y) \right\} \SL 0 \,,  \label{Eq:WTEA}
\end{align}
where we performed integration by parts. Meanwhile, multiplying 
$e^{-i \sbm{k}_L' \cdot \sbm{x}}$ on Eq.~(\ref{Exp:WT0}) and integrating over $x^\mu$, we obtain
\begin{align}
 & \int d^{\dd +1} x e^{-i \sbm{k}_L' \cdot \sbm{x}}   \int d^{\dd +1} y
 e^{-i \sbm{k}_L \cdot \sbm{y}}  \left\{ W^{(2)}_{\delta g_{\pm} \zeta_{\pm}}(x,\, y) +
 W^{(2)}_{\delta g_{\pm} \zeta_{\mp}}(x,\, y) \right\} = 0 \,.   \label{Eq:WTEAct}
\end{align}

By adding the left hand side of Eq.~(\ref{Eq:WTEA}) multiplied by a constant parameter $- s$ and
Eq.~(\ref{Eq:WTEAct}) with $\delta g_\pm=\zeta_\pm$ multiplied by $- s^2/2$, the linear and the
quadratic terms in the effective action can be given by 
\begin{align}
 &  i \Seffd{1}[\delta g_+,\, \delta g_-]  +  i \Seffd{2}[\delta g_+,\,
 \delta g_-]
 \cr
 & =  \sum_{a = \pm} \int
 d^{\dd +1} x\, \delta g_{a,\, s}(x) W^{(1)}_{\delta g_a} (x) \cr
 & \qquad +  \frac{1}{2!} \!\sum_{a_1, a_2 = \pm}\!  
 \int d^{\dd +1} x_1 \!\int\! d^{\dd +1} x_2 \, \delta g_{a_1,\,
 s}(x_1) \delta \tilde{g}_{a_2,\,s}(x_2) W^{(2)}_{\delta g_{a_1}
 \delta \tilde{g}_{a_2}}(x_1,\,x_2) \cr
 & \qquad + {\cal O}(\delta g^3,\, \delta g^2 s,\, \delta g s^2, s^3) \,, \label{Expn:Seffdd}
\end{align}
where $\delta g_s$ denote the metric perturbations $\delta g$ after the
inhomogeneous dilatation. For notational brevity, here we used the same notation as those for the
global dilatation, given in Eqs.~(\ref{Exp:transzeta}) and
(\ref{Exp:transN}). Here, each $\delta g_{i, a}~(i=1,\, 2)$ sums over 
$\delta N_{a, s}$, $N_{i, a, s}$, $\zeta_{a, s}$, and $\gamma_{ij, a, s}$. 
In deriving Eq.~(\ref{Expn:Seffdd}),
we used 
\begin{align}
 & W^{(2)}_{\delta g_{a_1} \delta \tilde{g}_{a_2}}(x_1,\, x_2) =
 W^{(2)}_{\delta \tilde{g}_{a_2} \delta g_{a_1}}(x_2,\,
 x_1)\,.
\end{align}
The first term in Eq.~(\ref{Eq:WTEA}) changes the argument of the metric perturbations in
the linear term of the metric perturbations in Eq.~(\ref{Expn:Seffdd}). We also changed the arguments of the
quadratic terms, taking into account that the modification appears only
in higher orders of perturbation. The tadpole contributions, which are the terms in the second
line of Eq.~(\ref{Expn:Seffdd}) should vanish by using the the
background equation of motion. (See Ref.~\cite{TY_conservation} for a
discussion about the heavy scalar field.)

Equation (\ref{Expn:Seffdd}) shows that with the use of the consistency
relation, $\delta g_\alpha(x)$ in $\Seff'$ can be replaced with $\delta g_{\alpha,\,s}(x)$. 
Since the rest of the effective action, $S_{ad}$, is simply
the classical action for the single field model, it also should be invariant
under this replacement. Therefore, when the locality condition (\ref{Cond:hard})
holds, the total effective action $S_{\rm eff}$ preserves the invariance
under the inhomogeneous dilatation with the suitable fall-off , i.e.,
under the change of $\delta g_\alpha$ to $\delta g_{s,\alpha}$. Since
Eqs.~(\ref{Eq:WTEA}) and (\ref{Eq:WTEAct}) also hold for the
soft modes not only for the homogeneous mode with $\bm{k}=0$, we can
shift $\zeta(x)$ by an inhomogeneous but time-independent 
function instead of the homogeneous constant parameter. The invariance of the effective action, which also includes the
radiative corrections from the non-zero spin massive fields, under the
shift of the soft mode $\zeta_{\sbm{k}_L}$ 
directly implies the existence of the constant solution for
$\zeta$\footnote{The lapse function and the shift vector, included in
the effective action (\ref{Expn:Seffdd}), can be eliminated by solving the Hamiltonian and momentum
constraint equations and expressing them in terms of $\zeta_s$ as in the
single field model~\cite{Maldacena02}. Since the constraint equations
for $\delta g_s$ are simply given by replacing $\delta g$ with $\delta g_s$ in the
constraint equations for $\delta g$, the effective action obtained after
eliminating these Lagrange multiplies obviously preserves the invariance
under the replacement of $\zeta$ with $\zeta_s$.}.

We assumed the locality condition (\ref{Cond:hard}) to derive the consistency
relation, which was used to show the invariance under the replacement of $\zeta$ with $\zeta_s$ in the
effective action. As was argued in Sec.~\ref{SSSec:NandNP}, the validity
of Eq.~(\ref{Cond:hard}) requires the linear perturbation of the soft mode
$\zeta_{\sbm{k}_L}$ to be time independent. For that, the decaying mode
of $\zeta_{\sbm{k}_L}$ should die off sufficiently fast after the Hubble
crossing. This happens when the background trajectory is on
an attractor, e.g., when the heavy fields do not alter this nature 
of the background
classical trajectory. As far as the radiative
corrections of the heavy fields $\bch$ do not turn the ``decaying'' mode
of $\zeta_{\sbm{k}_L}$ into a 
growing mode, the existing constant solution should be the dominant
solution of $\zeta$ in the large scale limit. This should be the case,
when the radiative corrections of the heavy fields remain perturbative. 
In Ref.~\cite{TY_conservation}, we listed the condition for
the conservation of $\zeta$ in the presence of the radiative corrections
of a massive (scalar) field as
\begin{itemize}
 \item The radiative corrections of the heavy field are perturbatively small.
 \item The background trajectory is on an attractor.
 \item The quantum system preserves the dilatation invariance. 
\end{itemize}
Now the second and third conditions are rephrased by the single condition:
\begin{itemize}
 \item The locality condition (\ref{Cond:hard}) is satisfied.
\end{itemize}
This guarantees the presence of the constant solution also for the soft
modes $\zeta_{\sbm{k}_L}$ with $\bm{k}_L \neq 0$, not only for the
homogeneous mode. The current argument also can apply to massive
fields with arbitrary spins, including higher spin fields.

\section{IR divergences of inflationary correlators} \label{Sec:IR}
In the previous two sections, we showed that the locality condition
(\ref{Cond:hard}) leads to the consistency relation and the conservation
of $\zeta$. In this section, as another related subject, we show that
the condition (\ref{Cond:hard}) also plays an important role for the
cancellation of the IR divergent contributions.

\subsection{Overview of IR divergence problem}
It is widely known that the loop corrections of a massless perturbation mode 
such as the curvature perturbation $\zeta$ can yield various IR
enhancements. (See, e.g., Refs.~\cite{Tsamis:1993ub, TW96, Sloth:2006az, Sloth:2006nu, Seery:2007we,
	Seery:2007wf, Urakawa:2008rb, Kitamoto:2012vj, Kitamoto:2013rea}.) In this section,
following Ref.~\cite{IRreview}, we briefly summarize the IR
enhancements. When we perform the perturbative expansion in terms of the
interaction picture field $\zeta_I$, an interaction vertex which
includes $\zeta_I$ without derivative can yield the radiative
correction which is proportional to $\langle \zeta_I^2 \rangle$. The
super Hubble modes with $k \alt aH$ contribute to $\langle \zeta_I^2 \rangle$ as
$\int_0^{aH} d^3k/k^3$, yielding the logarithmic enhancement. We
distinguish the divergent contribution due to the modes $ 0 \leq k \leq k_c$
(IRdiv) from the convergent but secularly growing one ($\propto \ln a$) due to the modes
$k_c \leq k \leq aH$ (IRsec). Here, $k_c$ denotes an IR cutoff. 
The IRsec, which cannot be removed by introducing the comoving IR
cutoff, originates from the accumulation of the super Hubble modes. 
The accumulation of the super Hubble modes enhances the time integral
at each interaction vertex (SG), introducing another secularly growing term
proportional to $\ln a$ or increasing the power of $\ln a$ included in
the integrand. These IR enhancements also can be introduced by the soft
graviton.

\subsection{Cancellation of IR divergence and the locality condition}
In the series of papers \cite{IRgauge_L, IRgauge,SRV2, IRreview, IRsingle, IRmulti, IRgauge_multi,SRV1,  SRVGW}, we showed that the IR
enhancements, i.e., IRdiv, IRsec, and SG, are due to the influences from
the outside of the observable region. What we
observe in cosmological measurements corresponds to a 
quantity evaluated in a completely fixed gauge. Fixing the gauge
conditions eliminates the influence of the gauge degrees of freedom. 
However, it is not possible to determine the gauge condition outside the
observable region, even if we completely specify the way of
observation. Therefore, unless the causality is manifestly ensured as in
the harmonic gauge, the degrees of freedom 
outside the observable region can affect the boundary
conditions of the observable region. As was argued in
Refs.~\cite{IRgauge_L, IRgauge}, changing the boundary conditions
corresponds to changing the spatial coordinates in the local observable
region. Such spatial coordinate transformations are the large
gauge transformations. We dubbed a variable which is independent of
those boundary conditions of the local observable region as a genuine
{\it gauge} invariant variable.

In Refs.~\cite{IRgauge_L,
IRgauge, SRV2, IRreview, IRsingle, SRV1, SRVGW}, we showed that all the IR
enhancements are cancelled out, when we calculate a correlation function
of a genuine {\it gauge} invariant variable for a specific initial state
such as the adiabatic vacuum. In Ref.~\cite{SRV1}, we argued that
whether the IR enhancements in the correlation function of the genuine
{\it gauge} invariant operator disappear or not depends on the choice of
the initial states. 
In this subsection, scrutinizing the condition on the quantum state to
ensure the cancellation of the IR enhancements, we show that when the locality condition
(\ref{Cond:hard}) is satisfied, the IRdiv, which comes from the momentum integral, is cancelled in the correlation
functions of the invariant variable under the large gauge
transformations. In the following, we denote a genuine
gauge invariant variable as ${^g\!R}(x)$ without specifying it. (One way to
construct ${^gR}(x)$ was discussed in Refs.~\cite{IRgauge_L, IRgauge}.)

One important property of ${^g\!R}$ is being constructed only by local
quantities such that commute with the soft modes and
remains invariant under the inhomogeneous dilatation, i.e.,
\begin{align}
 & {^g\!R}(x) =  e^{i Q^W_\zeta(\sbm{k}_L)} {^g\!R}(x) e^{- i
 Q^W_\zeta(\sbm{k}_L)} \,. \label{Exp:gRinv}
\end{align}
As in Eq.~(\ref{Exp:psiz0}), but focusing on a single soft mode, 
we expand the correlation function of
${^g\!R}$ as
\begin{align}
 & \langle \Psi\, | {^g\!R}(x_1) \cdots {^g\!R}(x_n) |\, \Psi \rangle
 \cr
 & = \int d \tilde{\zeta}^c_{\sbm{k}_L}
 |\psi(\tilde{\zeta}^c_{\sbm{k}_L})|^2 \,\,\,
 {}_{\tilde{\zeta}^c_{\sbm{k}_L}}\!\! \langle \Psi\,|  {^g\!R}(x_1) \cdots
 {^g\!R}(x_n) |\, \Psi \rangle_{\tilde{\zeta}^c_{\sbm{k}_L}}\,,  \label{Exp:exp}
\end{align}
where we noted that the soft mode $\tilde{\zeta}_{\sbm{k}_L}$ commutes
with the genuine gauge invariant variable. When the ``locality''
condition holds, since ${^g\!R}$ commutes with $Q^W_\zeta(\bm{k}_L)$
we obtain
\begin{align}
0 &=  {}_{\tilde{\zeta}^c_{\sbm{k}_L}}\!\! \langle \Psi\,| \left[ i
 Q^W_\zeta(\bm{k}_L),\,  {^g\!R}(x_1) \cdots  {^g\!R}(x_n) \right] |\,
 \Psi \rangle_{\tilde{\zeta}^c_{\sbm{k}_L}} \cr
  & = \frac{\partial}{\partial \tilde{\zeta}^c_{\sbm{k}_L}} \,\,
 {}_{\tilde{\zeta}^c_{\sbm{k}_L}}\!\! \langle \Psi\,| {^g\!R}(x_1)
 \cdots  {^g\!R}(x_n) |\, \Psi \rangle_{\tilde{\zeta}^c_{\sbm{k}_L}} \,.
\label{IRcancel}
\end{align}
Since the correlator in Eq.~(\ref{IRcancel}) is
independent of the soft mode $\tilde{\zeta}^c_{\sbm{k}_L}$, it commutes
with the integral over $\tilde{\zeta}^c_{\sbm{k}_L}$. Then, the
divergent integral 
$\int d \tilde{\zeta}^c_{\sbm{k}_L}|\psi(\tilde{\zeta}^c_{\sbm{k}_L})|^2$
in Eq.~(\ref{Exp:exp}) simply becomes the normalization factor, which
should be canceled in computing connected diagrams. Here, we picked up a certain wavenumber
$\bm{k}_L$. However, repeating the same procedure for the whole soft
modes, we find that all the soft modes which correlate with the
hard modes are canceled out. This cancellation yields a suppression of
the soft modes which interact with the hard modes and ensures the
absence of the IRdiv in the correlation function of ${^g\!R}$. In this way, we find that while the quantum state $|\, \Psi \rangle$ is
not invariant under the inhomogeneous dilatation, which inserts the soft mode
$\zeta_{\sbm{k}_L}$, the correlation function of ${^g\!R}$ for
the quantum state $|\, \Psi \rangle$ is insensitive to the insertion. 
In Ref.~\cite{IRNG}, this cancellation of the correlation between the
soft modes and the hard modes was presented by considering the squeezed bi-spectrum.

Here, let us further discuss the relation between the genuine 
{\it gauge} invariance and the absence of the IRdiv. Changing the
boundary condition at the edge of the observable region, we 
can alter the spatial average of the curvature perturbation
$\zeta$~\cite{IRsingle, IRgauge_L, IRgauge}. This can be expressed as
the dilatation whose constant parameter $s$ is given by
the spatial average of $\zeta$ in the observable region. This dilatation
changes the constant part of all the modes with $k_L \alt 1/L_O$, where $L_O$ is the size of the observable region, not
only the homogeneous mode with $\bm{k}=0$. Therefore, the genuine 
{\it gauge} invariance requires being insensitive to the excitation of
the constant soft modes $\zeta_{\sbm{k}_L}$. As we argued in
Sec.~\ref{SSSec:NandNP} (see around Eq.~(\ref{Changezeta})), quantum states 
which satisfy the
locality condition do not preserve the genuine {\it gauge}
invariance in the sense that it is not insensitive to the insertion of 
the constant part of the soft modes $\zeta_{\sbm{k}_L}$. What preserves the genuine {\it gauge} invariance is
the correlation function of the genuine 
{\it gauge} invariant operator evaluated for such 
quantum states\footnote{
Although the locality condition is necessary condition for the absence of 
IRdiv, this does not immediately imply that the locality condition is a requirement 
for the quantum state of the whole universe. When we discuss observables for a local observer, 
it would be allowed to trace out the degrees of freedom which 
the observer cannot see. 
After tracing out these degrees of freedom, the density matrix of the universe will be 
block diagonalized with a good precision. Then, the observables will 
correspond to the expectation values just for one of the 
blocks in the density matrix. In this sense, 
the actual observables are likely to be quite different from 
the simple expectation values for a given wave functional of the 
whole universe.}. While the relation between the
dilatation invariance and the cancellation of the IRdiv has been
discussed in a number of literatures, e.g., in Refs.~\cite{IRgauge_L,
IRgauge, SRV2, IRsingle, SRV1, BGHNT10, GHT11, GS10, GS11, SZ1203}, 
this aspect has not been clearly described elsewhere.

By contrast, since the curvature perturbation $\zeta_{\sbm{k}}$ is not
a genuine {\it gauge} invariant operator, the correlation functions which
includes the operator $\zeta_{\sbm{k}}$ suffers from the IRdiv (and also
IRsec and SG). Because of that, the 
correlation functions
$\langle \zeta_{\sbm{k}_L} O_{\sbm{k}_{S1}} \cdots O_{\sbm{k}_{Sn}} \rangle$
diverge due to the accumulation of the soft modes. Here, $O_{\sbm{k}_{S}}$ is either $\zeta_{\sbm{k}_S}$
or ${\cal O}y_{\sbm{k}_S}$. In order to make these correlation functions 
finite, we need to somehow introduce IR regularization. 
However, recall that a naive introduction of the IR
cutoff violates the dilatation invariance, which was the starting point
of the discussion about the consistency relation. Therefore, to be precise, we should not understand
the consistency relation as the relation between the correlation functions for the
Heisenberg operators  
$\zeta_{\sbm{k}_L} O_{\sbm{k}_{S1}} \cdots O_{\sbm{k}_{Sn}}$ and
$O_{\sbm{k}_{S1}} \cdots O_{\sbm{k}_{Sn}}$, since they are not well-defined. 
Instead, the consistency relation we discussed in this paper should be
understood as the relation between the ``correlation function'' for the hard
modes $O_{\sbm{k}_{S1}} \cdots O_{\sbm{k}_{Sn}}$ without any
propagation of the soft modes and the one with the additional insertion
of the free soft mode as an external leg. Then, both of them do not
contain the loop corrections of the soft modes, which can lead to the IR
enhancements. This is the reason why we needed to employ the
approximation $\SL$ in deriving the consistency relation.

\section{Relevance and irrelevance of soft graviton insertion}  \label{Sec:GW}
The curvature perturbation $\zeta$ and the graviton
$\gamma_{ij}$ are both massless fields and they have similar IR
behaviours. In this section, we briefly show that the discussion about
the consistency relation and the IR divergence for the graviton proceed
almost in parallel to those for the curvature perturbation
$\zeta$. For this purpose, we consider the shear transformation, which
is a large gauge transformation: 
\begin{align}
 & x^i\, \to \, \tilde{x}^i \equiv \left[ e^{\frac{S}{2}} \right]^i\!_j\, x^j \,, \label{Exp:LGT}
\end{align}
where $S_{ij}$ is a constant symmetric and traceless tensor $S_{ij}$. Under this large gauge
transformation, the spatial metric transforms as
\begin{align}
 & \left[ e^{\tilde{\gamma}(t,\, \tilde{\sbm{x}})} \right]_{ij} = \left[ e^{- \frac{S}{2}} \right]_i\!^k
  \left[ e^{- \frac{S}{2}} \right]_j\!^l \,  \left[
 e^{\gamma(t,\, \sbm{x})} \right]_{kl}\,. 
\end{align}
At the linear perturbation, $\gamma_{ij}$ is shifted as 
$\tilde{\gamma}_{ij} = \gamma_{ij} -S_{ij}$. The classical action is invariant
under the large gauge transformation (\ref{Exp:LGT}).

Similar to the dilatation, we define the Noether charge for
the large gauge transformation (\ref{Exp:LGT}) as
\begin{align}
 & Q_\gamma \equiv \frac{1}{2} \int d^3 \bm{x} \left[  \Delta_S \gamma_{ij}(t,\,
 \bm{x}) \pi_\gamma^{ij} (t,\,  \bm{x}) + \pi_\gamma^{ij}(t, \, \bm{x}) \Delta_S
 \gamma_{ij}(t,\, \bm{x}) \right]\,,  \label{Exp:QGWP}
\end{align}
where $\pi_{\gamma}^{ij}$ denotes the conjugate momentum of
$\gamma_{ij}$ and
\begin{align}
 & \Delta_S \gamma_{ij}(t,\,\bm{x}) \equiv \tilde{\gamma}_{ij} (t,\,
 \bm{x}) - \gamma_{ij} (t,\, \bm{x})\,. 
\end{align}
The invariance of the quantum state under this transformation requires
\begin{align}
 & Q_\gamma |\, \Psi \rangle =0\,. 
\end{align}
Repeating a similar argument, we find that the invariance is preserved,
when the following conditions
\begin{align}
 & \frac{\partial}{ \partial \bar{\gamma}^c} |\psi(\bar{\gamma}^c)| = 0\,, \label{Cond:homg0}
 \\[2pt]
 & i Q_\gamma |\, \Psi \rangle_{\bar{\gamma}^c} =  S_{ij} \frac{\partial}{\partial
 \bar{\gamma}_{ij}^c} |\, \Psi \rangle_{\bar{\gamma}^c} \,, \label{Cond:homg}
\end{align}
are satisfied. Here, $\bar{\gamma}_{ij}$ denotes the homogeneous mode of
$\gamma_{ij}$ and $ |\, \Psi \rangle_{\bar{\gamma}^c}$ denotes the
projected quantum state into the eigenstate of $\bar{\gamma}^c_{ij}$.

Inserting the window function $W(\bm{x})$ into the integrand of the Noether charge
$Q_\gamma$ and performing the Fourier transformation, we define $Q^W_\gamma(\bm{k}_L)$, 
which inserts the soft graviton $\gamma_{ij, \sbm{k}_L}$. Again, we find
that when the condition (\ref{Cond:homg}) can be extended to the soft
modes with $\bm{k}_L \neq 0$, i.e.,
\begin{align}
 &  i Q_\gamma^W(\bm{k}_L) |\, \Psi
 \rangle_{\tilde{\gamma}^c_{- \sbm{k}_L}} =  (2 \pi L_c)^3 S_{ij\, - \sbm{k}_L}
  \frac{\partial}{\partial
 \tilde{\gamma}_{ij\, - \sbm{k}_L}^c} |\, \Psi
 \rangle_{\tilde{\gamma}^c_{- \sbm{k}_L}}\, \label{Cond:hardgamma}
\end{align}
with $S_{ij\, \sbm{k}_L} \equiv \hat{W}(\bm{k}_L) S_{ij}$,
we obtain the consistency relation which relates the correlation functions for
the hard modes and those with single insertion of the Gaussian soft
graviton (see, e.g., Refs.~\cite{Maldacena02, Bordin:2016ruc}). The quantum state $|\, \Psi \rangle$ changes due to the
single insertion of the soft graviton, i.e.,
\begin{align}
 & Q_\gamma^W(\bm{k}_L) |\, \Psi \rangle \neq 0\,, \label{Changegamma}
\end{align}
because the wave function is not completely flat in the direction
of the soft graviton $\gamma_{ij\, \sbm{k}_L}$ with $\bm{k}_L \neq 0$
in contrast to the homogeneous mode, whose wave function is completely
flat, satisfying Eq.~(\ref{Cond:homg0}). Let us emphasize again that this is not a spontaneous symmetry breaking,
because the large gauge transformation (\ref{Exp:LGT}) with an
inhomogeneous $S_{ij}$ is not a symmetry of the classical action.

The correlation function of the genuine {\it gauge} invariant operator
${^gR}$ is insensitive to the insertion of the soft graviton. This
ensures the absence of the IRdiv due to the soft graviton in the
correlation functions of ${^gR}$ evaluated for $|\, \Psi \rangle$. By contrast, when
we evaluate a correlation function for an operator which does not
preserve the genuine {\it gauge} invariance, the insertion of the soft
graviton changes the correlation function and this can lead to a break
down of the perturbative expansion~\cite{Ferreira:2016hee, Ferreira:2017ogo}. 
(See also Ref.~\cite{GT07}.)

\section{Concluding remarks}  \label{Sec:Conclusion}
The relation among the large gauge transformations, the consistency
relations for the soft modes of $\zeta$ and $\gamma_{ij}$, their
conservation in time, and the IR enhancements has been
discussed in a number of literatures. However, as far as we understand, this relation was not
fully clarified and it was sometimes understood in a misleading manner. The purpose
of this paper is to sharpen the argument about the relation among these  
four. The
invariance of the quantum state $|\, \Psi \rangle$ under the dilatation
and shear transformations can be preserved, when the following two
conditions are fulfilled. First, the amplitude of the wave function $\psi$ should be flat
towards the directions for the homogeneous modes of $\zeta$ and $\gamma_{ij}$.
Second, operating the Noether charges $Q_\zeta$ and $Q_\gamma$ on 
the quantum state of hard modes is equivalent to additively and time-independently 
shifting the homogeneous modes $\bar{\zeta}$ and $\bar{\gamma}_{ij}$,
which interact with the hard modes as described in Eqs.~(\ref{Cond:inh0})
and (\ref{Cond:homg}).

The invariance under these large gauge transformations, just itself,
leads to neither the consistency relation nor the absence of
IRdiv. The additional conditions (\ref{Cond:inh0}) and
(\ref{Cond:homg}) are the non-trivial extensions 
to those which describe the insertion of the soft modes
$\bm{k}_L (\neq 0)$, which are not
always satisfied. These conditions can be interpreted as the locality condition, which states
that the inhomogeneous dilatation and shear transformations only change
the values of $\zeta$ and $\gamma_{ij}$ within each local
universe. Since the wave function is not flat in the directions of the 
soft modes for the curvature perturbation and the graviton, the quantum
state $|\, \Psi \rangle$ changes due to the insertion of these soft
modes. When the locality conditions (\ref{Cond:hard}) and (\ref{Cond:hardgamma})
are satisfied, the influence of these soft modes are described by the
well-known consistency relations under the Gaussian approximation of the
wave function for the soft modes. This argument also applies in 
deriving the consistency relation for massive
fields. We also showed that the locality condition
(\ref{Cond:hard}) implies the conservation of $\zeta$ in the soft limit
within the perturbation theory. The same argument also applies to
$\gamma_{ij}$.

The final issue is the IR enhancements due to the soft modes of $\zeta$
and $\gamma_{ij}$. In contrast to the correlation functions for $\zeta$
and $\gamma_{ij}$, which are not genuinely {\it gauge} invariant, the
correlation functions for a genuine {\it gauge} invariant operator
remain invariant under the
insertion of the soft modes for the curvature perturbation $\zeta$ and
the graviton $\gamma_{ij}$, when the locality conditions hold. This ensures
the absence of the IRdiv due to the soft modes of $\zeta$ and $\gamma_{ij}$.
In this paper, we did not discuss the IRsec and the SG, which yields the secular
growth. When the locality conditions hold at each time slicing, repeating the same argument
as in Refs.~\cite{SRV2, SRVGW}, we can show the absence of the IRsec and
SG. In fact, this is the case when we choose the adiabatic vacuum (or the
Euclidean vacuum) as the initial state of the universe.

Recently, the relation among the asymptotic symmetry, the soft
theorem, and the IR divergence was discussed about gauge theories in
asymptotically flat spacetime~\cite{Strominger:2013jfa, He:2014laa, Strominger:2014pwa,Kapec:2017tkm}. In Ref.~\cite{Strominger:2013jfa}, it was shown
that the Weinberg's soft theorem for the soft photons and gravitons can be
derived as the Ward-Takahashi identities for the asymptotic symmetry. (For
a recent review, see Ref.~\cite{Strominger:2017zoo}.) In Ref.~\cite{Kapec:2017tkm}, the relation
between the asymptotic symmetry and the IR divergence of the QED was
discussed. About the IR divergence in QED, Faddeev and Kulish showed that the IR
finiteness can be guaranteed, when we consider the dressed charged
particles by soft photon clouds. (See also Refs.~\cite{Chung:1965zza, Zwanziger:1973if}.) While there is a qualitative difference
between the in-out formalism in QED and the in-in formalism in
cosmology, our genuine {\it gauge} invariant operator ${^gR}$, whose
correlators can be IR finite with an appropriate choice of the initial
state, also dresses the clouds of the soft $\zeta$ and $\gamma_{ij}$ as
external legs.

At first glance, the IR structures for the gauge fields in the
asymptotically flat spacetime have a certain similarity to those for
the primordial perturbations $\zeta$ and $\gamma_{ij}$. However, a
closer look may also reveal some differences in these two cases. For the
gauge theories in the asymptotically flat spacetime, we are interested
in the transition at the asymptotic infinity between before and after
the propagations of the soft photons and gravitons. On the other hand,
for the primordial perturbations, the asymptotic infinity is the
spatial infinity and is out of reach in a causal evolution. Instead, what
we are interested in is a locally defined quantity like an actually
observable quantity. As is summarized above, in order to derive the consistency relation,
the soft theorem for $\zeta$ and $\gamma_{ij}$, we need to assume the
locality condition, which is not trivially satisfied even in a Diff
invariant theory, at least in the current gauge choice (the Maldacena
gauge~\cite{Maldacena02}). This gives a qualitative difference from the
gauge theories in the asymptotically flat spacetime, which trivially
satisfies the locality.

\acknowledgments
We would like to thank M.~Mirbabayi for a fruitful conversation. T.~T. was also
supported in part by MEXT Grant-in-Aid for Scientific Research on Innovative
Areas, Nos. 17H06357, 17H06358, 24103001 and 24103006,
and by Grant-in-Aid for Scientific Research (A)  No. 15H02087. Y.~U. is supported by JSPS Grant-in-Aid for Research Activity Start-up under Contract No.~26887018, Grant-in-Aid for Young Scientists (B) under Contract No.~16K17689, 
and Grant-in-Aid for Scientific Research on Innovative Areas under Contract No.~16H01095. This research is
supported in part by Building of Consortia for the Development of Human Resources in
Science and Technology and the National Science Foundation under Grant No. NSF
PHY11-25915.

\appendix

\section{Perturbative and non-perturbative definition of Noether charge} \label{Ap:NC}
%

In this section, we show that performing the dilatation and
performing the perturbative expansion are not commutable processes,
i.e., the dilatation transformation in the Heisenberg picture and the
one in the interaction picture lead to different expressions. In order
to see this, let us first consider the dilatation in the Heisenberg
picture with the use of the Noether charge 
$Q_\zeta$ as  
\begin{align}
 & \zeta_s(x) = e^{i Q_\zeta} \zeta(x) e^{- i Q_\zeta} = \zeta(x) +
 \Delta_s \zeta(x) \,.
\end{align}
Then, we perturbatively expand both $\zeta(x)$ and $\zeta_s(x)$, i.e., before and after
the dilatation transformation, following the standard procedure, as
\begin{align}
 & \zeta(t,\, \bm{x}) = U^{I \dagger}(t)\, \zeta^{I}(t,\, \bm{x})\,
 U^{I}(t)\,, \label{Exp:interactionS0}
\end{align}
and
\begin{align}
 & \zeta_s(t,\, \bm{x}) = U^{I \dagger}_s(t)\, \zeta_s^{I}(t,\, \bm{x})\,
 U^{I}_s (t)\,, \label{Exp:interactionS}
\end{align}
where $U^{I}$ and $U^{I}_s$ denote the unitary operators which
relate the Heisenberg and interaction picture fields for before and
after the dilatation transformation. Using the interaction Hamiltonian
$H^{I} \equiv H - H_0$, the unitary operator is given by
\begin{align}
 & U^I(t) \equiv T e^{- i \int^t d t' H^I}\,, \qquad
 U_s^I(t) \equiv T e^{- i \int^t d t' H_s^I} \,. 
\end{align} 
Since the free Hamiltonian $H_0$ changes due to the dilatation
transformation while the total Hamiltonian does not, the interaction
Hamiltonian $H^I$ and $U^I$ also change through the dilatation, i.e.,
\begin{align}
 & U^{I\, \dagger}_s (t) U^I(t) = 1 + {\cal O}(s)\,. \label{Exp:changeU}
\end{align}

Next, we show that the interaction picture field $\zeta_s^I(t,\, \bm{x})$,
which is related to $\zeta_s$ as in Eq.~(\ref{Exp:interactionS}), does
not coincide with
\begin{align}
 & \tilde{\zeta}^I_s(x) \equiv e^{i Q_s^I} \zeta_I(x) e^{- i Q_s^I} =
 \zeta_I(x) + \Delta \zeta_I(x)\,,  \label{Exp:tildezeta}
\end{align}
which is given by performing the dilatation transformation in the
interaction picture. Here, $\Delta \zeta_I$ is given by replacing the
Heisenberg fields with the interaction picture fields in $\Delta \zeta(x)$.  
Notice that $\tilde{\zeta}^I_s(x)$ is related to $\zeta_s(x)$ by the
unitary operator $U^I(t)$, i.e.,
\begin{align}
 & \tilde{\zeta}^I_s(x) = U^I (t) \zeta_s(t,\, \bm{x}) U^{I \dagger}(t)
\end{align}
while the standard perturbative prescription in the frame after the dilatation
transformation uses the interaction picture field given by
\begin{align}
 \zeta_s^I(t, \bm{x}) &= U_s^I (t) \zeta_s(t,\, \bm{x}) U_s^{I
 \dagger}(t) . 
\end{align}

As is shown in Eq.~(\ref{Exp:changeU}), since the unitary operator $U_I(t)$ changes
under the dilatation transformation, we obtain
\begin{align}
 & \tilde{\zeta}_s^I(x) - \zeta^I_s(x) = {\cal O}(s)\,. 
\end{align}
Therefore, the dilatation transformation in the interaction picture
(\ref{Exp:tildezeta}) does not give the interaction picture field
defined in the standard prescription of perturbation
theory after the dilatation transformation, i.e., Eq.~(\ref{Exp:interactionS}).
This discrepancy vanishes by sending $s$ to 0.

\section{Computing the effective action} \label{Sec:IFcomp}
In this Appendix, we derive the expression of $W^{(n)}_{\delta g_{a_1}
\cdots \delta g_{a_n}}(x_1,\, \cdots, \,x_n)$, defined in Eq.~(\ref{Def:tWn}). 
The linear term in the effective action is given by
\begin{align}
 i \Seffd{1} &=  \sum_{a= \pm} \int d^{\dd +1} x\,
  \delta g_a(x) W^{(1)}_{\delta g_\alpha} (x)
\,, 
\end{align}
where $W^{(1)}_{\delta g_\alpha}$ is given by the expectation value as
\begin{align}
 & W^{(1)}_{\delta g_+} (x) = - W^{(1)}_{\delta g_-} (x) =  \left\langle
 \frac{\delta i  S_\chi}{\delta g(x)}  \bigg|_{\delta g=0}  \right\rangle\,.
\end{align}
Next, we compute the quadratic terms in $\Seff'$. Taking
the second variation of $\Seff'$ with respect to $\delta g_+$, we obtain
\begin{align}
 W^{(2)}_{\delta g_+ \delta \tilde{g}_+}(x_1,\,x_2) & = i^2 \left\langle
 \frac{\delta S_\chi[\delta g_+,\, \bch_+]}{\delta g_+(x_1)}
 \bigg|_{\delta g_+=0} \frac{\delta S_\chi[\delta g_+,\, \bch_+]}{\delta
 \tilde{g}_+(x_2)} \bigg|_{\delta g_+=0}  \right\rangle_{\pm} \cr
 & \qquad \qquad \qquad \quad  + i \delta(x_1-x_2) \left\langle
 \frac{\delta^2 S_\chi[\zeta_+,\, \bch_+]}{\delta g_+(x_1) \delta \tilde{g}_+(x_1)}
 \bigg|_{\delta g_+=0} \right\rangle_{\pm}\,, \label{Exp:dS++}
\end{align}
where $\delta g$ and $\delta \tilde{g}$ are either $\delta N$, $N_i$,
$\zeta$, or $\gamma_{ij}$. Here, we introduced the expectation value: 
\begin{align}
 & \langle {\cal O}[\bch_+,\, \bch_-] \rangle_{\pm} \equiv \frac{\int D \bch_+
 \int D \bch_- \,{\cal O}[\bch_+,\, \bch_-] e^{i S_\chi[0,\, \sbch_+]- i
 S_\chi[0,\,\sbch_-]}}{\int D \bch_+ \int D \bch_- \,e^{i S_\chi[0,\,\sbch_+]- i
 S_\chi[0,\,\sbch_-]}} \,.
\end{align}
Since the action $S_\chi[\delta g_+,\, \bch_+]$ includes only local terms, the variation of
$S_\chi[\delta g_+,\, \bch_+]$ with respect to $\delta g_+(x_1)$ and
$\delta \tilde{g}_+(x_2)$ yields the delta function $\delta(x_1 -x_2)$
in Eq.~(\ref{Exp:dS++}). Similarly, the second variation of $\Seff'$ with respect to $\delta g_-$ is given by 
\begin{align}
  W^{(2)}_{\delta g_- \delta \tilde{g}_-}(x_1,\,x_2) & = i^2 \left\langle
 \frac{\delta S_\chi[\delta g_-,\, \bch_-]}{\delta g_-(x_1)}
 \bigg|_{\delta g_-=0} \frac{\delta S_\chi[\delta g_-,\, \bch_-]}{\delta
 \tilde{g}_-(x_2)} \bigg|_{\delta g_-=0}  \right\rangle_{\pm} \cr
 & \qquad \qquad \qquad \quad  - i \delta(x_1-x_2) \left\langle
 \frac{\delta^2 S_\chi[\delta g_-,\, \bch_-]}{\delta g_-(x_1) \delta \tilde{g}_-(x_1)}
 \bigg|_{\delta g_-=0} \right\rangle_{\pm}\,. \label{Exp:dS--}
\end{align} 
Taking the derivative with respect to both $\delta g_+$ and 
$\delta g_-$, we obtain
\begin{align}
  W^{(2)}_{\delta g_+ \delta \tilde{g}_-}(x_1,\,x_2) & = - i^2 \left\langle
 \frac{\delta S_\chi[\delta g_+,\, \bch_+]}{\delta g_+(x_1)}
 \bigg|_{\delta g_+=0} \frac{\delta S_\chi[\delta g_-,\, \bch_-]}{\delta
 \tilde{g}_-(x_2)} \bigg|_{\delta g_-=0}  \right\rangle_{\pm} \,, \label{Exp:dS+-}
\end{align}
and
\begin{align}
 W^{(2)}_{\delta g_- \delta \tilde{g}_+}(x_1,\,x_2)
 & = - i^2 \left\langle
 \frac{\delta S_\chi[\delta g_-,\, \bch_-]}{\delta g_-(x_1)}
 \bigg|_{\delta g_-=0} \frac{\delta S_\chi[\delta g_+,\, \bch_+]}{\delta
 \tilde{g}_+(x_2)} \bigg|_{\delta g_+=0}  \right\rangle_{\pm} \,. \label{Exp:dS-+}
\end{align}
When the interactions of $\bch$ are perturbatively suppressed, we can
compute the functions $W^{(2)}_{\delta g_{a_1} \delta \tilde{g}_{a_2}}(x_1,\, x_2)$ 
by expanding them in terms of the free propagators for $\bch$.

\end{document}